\documentclass[11pt]{article}
\usepackage{amsmath, amssymb}
\usepackage{latexsym}
\usepackage{a4,epsfig}
\usepackage{amsfonts}
\usepackage{graphicx}
\usepackage{amscd}
\usepackage{amsbsy}

\numberwithin{equation}{section}
\newcommand\eq[1] {(\ref{#1})}

\newtheorem{lemma}{Lemma}[section]
\newtheorem{theorem}{Theorem}[section]

\newcommand\proof{{\it Proof:}\quad}

\newcommand\qed{\ \rule[-0.2ex]{0.3em}{1.5ex}}
\newcommand{\bfm}[1]{\mbox{\boldmath ${#1}$}}

\newcommand{\beqa}{\begin{eqnarray}}
\newcommand{\eeqa}[1]{\label{#1}\end{eqnarray}}
\newcommand{\beq}{\begin{equation}}
\newcommand{\eeq}[1]{\label{#1}\end{equation}}
\newcommand{\Grad}{\nabla}
\newcommand{\Div}{\nabla \cdot}
\newcommand{\Curl}{\nabla \times}

\newcommand{\Real}{\mathop{\rm Re}\nolimits}
\newcommand{\Imag}{\mathop{\rm Im}\nolimits}
\newcommand{\Tr}{\mathop{\rm Tr}\nolimits}

\newcommand{\lang}{\langle}
\newcommand{\rang}{\rangle}
\newcommand{\Md}{\partial}

\newcommand{\nm}{\noalign{\smallskip}}

\newcommand{\Ga}{\alpha}
\newcommand{\Gb}{\beta}
\newcommand{\Gd}{\delta}

\newcommand{\Gg}{\gamma}

\newcommand{\Gl}{\lambda}

\newcommand{\Gm}{\mu}

\newcommand{\Gs}{\sigma}

\newcommand{\Go}{\omega}

\newcommand{\GG}{\Gamma}
\newcommand{\GL}{\Lambda}

\newcommand{\GO}{\Omega}

\newcommand{\GY}{\Psi}

\newcommand{\BGs}{\bfm\sigma}

\def\Ba{{\bf a}}

\def\Be{{\bf e}}

\def\Bj{{\bf j}}

\def\Bn{{\bf n}}

\def\Bt{{\bf t}}

\def\Bv{{\bf v}}

\def\Bx{{\bf x}}
\def\By{{\bf y}}
\def\Bz{{\bf z}}

\def\BE{{\bf E}}
\def\BF{{\bf F}}

\def\BJ{{\bf J}}
\def\BK{{\bf K}}

\def\BU{{\bf U}}

\def \dis {\displaystyle}
\def \ep {\varepsilon}
\newcommand{\underj}{\underline{\Bj}}
\def \p {\partial}
\def \Om {\Omega}
\def \la {\lambda}
\def \ph {\varphi}
\def \RR {{\mathbb R}}

\def \ba {\begin{array}}
\def \ea {\end{array}}
\newtheorem {Thm} {Theorem} [section]

\newtheorem {Adef} [Thm] {Definition}

\newtheorem {Arem} [Thm] {Remark}

\newtheorem {Aexa} [Thm] {Example}
\newenvironment {Exa} {\begin{Aexa}\rm}{\end{Aexa}}
\newtheorem {Anot} [Thm] {Notation}

\def \refe #1.{(\ref{#1})}
\def \reff #1.{figure~\ref{#1}}
\def \refs #1.{section~\ref{#1}}
\def \refss #1.{subsection~\ref{#1}}
\def \refD #1.{Definition~\ref{#1}}
\def \refT #1.{Theorem~\ref{#1}}
\def \refL #1.{Lemma~\ref{#1}}
\def \refC #1.{Corollary~\ref{#1}}
\def \refP #1.{Proposition~\ref{#1}}
\def \refR #1.{Remark~\ref{#1}}
\def \refE #1.{Example~\ref{#1}}
\def \refN #1.{Notation~\ref{#1}}

\usepackage{graphics}
\usepackage{amssymb}
\begin{document}
\vspace{-1in}
\title{Sharp bounds on the volume fractions of two materials in a two-dimensional body from electrical boundary measurements: the translation method}
\author{Hyeonbae Kang\\
\small{Department of Mathematics, Inha University, Incheon 402-751, Korea} \\
\small{(hbkang@inha.ac.kr)}\\
\\
Eunjoo Kim\\
\small{Institute of Mathematical Sciences and Department of Mathematics,} \\
\small{Ewha Womans University, Seoul 120-750, Korea} \\
\small{(kej@ewha.ac.kr)}\\
\\
Graeme Milton\\
\small{Department of Mathematics, University of Utah, Salt Lake City, UT 84112, USA} \\
\small{(milton@math.utah.edu)}}
\date{}
\maketitle
\begin{abstract}
We deal with the problem of estimating the volume of inclusions using a finite number of boundary measurements in electrical impedance tomography. We derive upper and lower bounds on the volume fractions of inclusions, or more generally two phase mixtures, using two boundary measurements in two dimensions. These bounds are optimal in the sense that they are attained by certain configurations with some boundary data. We derive the bounds using the translation method which uses classical variational principles with a null Lagrangian. We then obtain necessary conditions for the bounds to be attained and prove that these bounds are attained by inclusions inside which the field is uniform. When special boundary conditions are imposed the bounds reduce to those obtained by Milton and these in turn are shown here to reduce to those of Capdeboscq-Vogelius in the limit when the volume fraction tends to zero. The bounds of this paper, and those of Milton, work for inclusions of arbitrary volume fractions. We then perform some numerical experiments to demonstrate how good these bounds are.
\end{abstract}
\vskip2mm

\noindent {\small Mathematics subject classification(MSC2000): Primary  35R30; Secondary 35A15}


\noindent
\vskip2mm

\section{Introduction}

\setcounter{equation}{0}

One of the central problems of the theory and practice of electrical impedance tomography is the problem of estimating the volume of the inclusions in terms of boundary measurements, either voltage measurements when currents are applied around the boundary of the body or current measurements when voltages are applied. The problem can described in rigorous terms as follows: Let $D$ be an inclusion inside a body $\Om$, and suppose that the conductivities of $D$ and $\Om \setminus D$ are $\Gs_1$ and $\Gs_2$ ($\Gs_1 \neq \Gs_2$), respectively. Let $\Gs= \Gs_1 \chi(D) + \Gs_2 \chi(\Om \setminus D)$ where $\chi(D)$ is the characteristic function of $D$ and the potential $V$ be the solution to
 \beq
 \left\{
 \begin{array}{l}
 \Grad \cdot \Gs \Grad V=0  \quad  \mbox{in } \Om, \\
 V=V^0 \quad \mbox{on } \p\Om
 \end{array}
 \right.
 \eeq{In1}
for some Dirichlet data (voltage) $V^0$ on $\p\Om$. Then the measurement of current (the Neumann data) is $q:= \Gs \frac{\p V}{\p \Bn}$ on $\p\Om$. (Throughout this paper $\frac{\p}{\p\Bn}$ denotes the normal derivative.) The problem is to estimate the volume $|D|$ of the inclusion using the boundary data $(V^0, q)$ for finitely many voltages, say $V^0=V_1^0, \ldots, V_n^0$. If the Neumann boundary condition $\Gs \frac{\p V}{\p \Bn} =q$ is prescribed on $\p\Om$ instead of the Dirichlet condition, then the measurement is $V^0:= V|_{\p\Om}$.

The purpose of this paper is to consider this problem and derive optimal upper and lower bounds for the volume fraction of inclusions in two dimensions. In fact, we deal with a more general situation where $\Om$ is a two phase mixture in which the phase 1 has conductivity $\Gs_1$ and the phase 2 has conductivity $\Gs_2$ ($\Gs_1 > \Gs_2$) so that the conductivity distribution $\Gs$ of $\Om$ is given by $\Gs(\Bx) = \Gs_1 \chi_1(\Bx) + \Gs_2 \chi_2(\Bx)$ where $\chi_j$ is the characteristic function of phase $j$ for $j=1,2$, {\it i.e.},
 \beq
 \chi_1 (\Bx)= 1- \chi_2(\Bx) = \left\{
 \begin{array}{l}
 1 \quad \mbox{in phase 1}, \\
 0 \quad \mbox{in phase 2}.
 \end{array}
 \right.
 \eeq{In2}
We derive optimal upper and lower bounds for the volume fraction $f_1$ of phase 1 ($f_1= \frac{1}{|\Om|} \int_{\Om} \chi_1 (\Bx)$) using boundary measurements corresponding to either a pair of Dirichlet data ($V_1^0$ and $V_2^0$) or a pair of Neumann data ($q_1$ and $q_2$) on $\Om$. The bounds are optimal in the sense that they are attained by some inclusions or configurations. The bounds can be easily computed from the boundary measurements. In fact, they are given by two quantities: the measurement (or response) matrix $A=(a_{ij})_{i,j=1,2}$ where
 \beq
 a_{ij} := \frac{1}{|\Om|} \int_{\p\Om} V_i^0 q_j
 \eeq{In3}
and
 \beq
 b_D := \frac{1}{|\Om|} \int_{\p\Om} V_1^0 \frac{\p V_2^0}{\p \Bt}
 \eeq{In4}
if the Dirichlet data are used. Here and throughout this paper, $\frac{\p}{\p \Bt}$ denotes the tangential derivative along $\p\Om$ in the positive orientation. If the Neumann data are used, then $b_D$ is replaced with
 \beq
 b_N := \frac{1}{|\Om|} \int_{\p\Om} q_1(\Bx) (\int_{\Bx_0}^{\Bx} q_2).
 \eeq{In5}
where the $\Bx_0\in\Md\GO$ and the last integral is on the surface $\Md\GO$.
See Theorem \ref{thm:LB1} and \ref{thm:UB1}.

Some significant results on the problem of estimating the volume of inclusion using boundary measurements are as follows. Kang-Seo-Sheen \cite{KSS97}, Alessandrini-Rosset \cite{AR98}, and Alessandrini-Rosset-Seo \cite{ARS00} obtained upper and lower bounds for the volume of the inclusion. However, their bounds involve constants which are not easy to determine, and hence it is not possible to compare them with the bounds of this paper. It is worth emphasizing that these results use only a single measurement. Another important result on volume estimation is that of Capdeboscq-Vogelius \cite{CV022, CV03}. They found, using the Lipton bounds on polarization tensors \cite{Lipton93}, upper and lower estimates for the volume of inclusions occupying a low volume fraction, which are optimal bounds in the asymptotic limit as the volume fraction tends to zero. Recently it was recognised by Milton \cite{milt11} that bounds on the response of two-phase periodic composites could be easily used to bound the multi-measurement response of two-phase bodies when special boundary conditions are imposed (see \eq{HS6} and \eq{HS10} below) and that these could be used in an inverse fashion to bound the volume fraction. As shown here those bounds coincide exactly with the Capdeboscq-Vogelius bounds in the asymptotic limit as the volume fraction tends to zero.

The bounds obtained in this paper allow for more general boundary conditions and we emphasize that they are optimal for any volume fraction. They reduce to those of Milton for the special boundary conditions, but have the advantage of being able to utilize the same set of measurements for
both the upper and lower volume fraction bounds. 
We derive the bounds using the
translation method which in its simplest form
is based on classical variational principles with null Lagrangians added, {\it i.e.}, non-linear functions of fields which may be integrated by parts and expressed in terms
of boundary measurements. The translation method,
developed by Murat and Tartar \cite{tar79,tar85,mutar85} and independently by Lurie
and Cherkaev \cite{lucherk82,lucherk84}, is a powerful method for deriving bounds on effective tensors of composites. As shown by Murat and Tartar it can be extended using the method of compensated compactness
to allow for functions more general than null Lagrangians, namely quasiconvex functions.
It is reviewed in the books \cite{cherk,milton,allaire,tartar}. The use of classical variational principles to determine information about the conductivity distribution inside a body from
electrical impedance measurements was pioneered by Kohn and Berryman \cite{kohnber}.

We continue our investigation by looking for necessary and sufficient conditions for the bounds to be attained. These are the exact analogs of the condition found by Grabovsky \cite{grab} for attainability
of the translation bounds for composites. (See also section 25.6 of \cite{milton}.). It turns out that the upper bound is attained if and only if the field in phase 1 is uniform and the lower bound is attained if and only if the field in phase 2 is uniform. It means that if phase 1 is an inclusion, the upper bound is attained if the field inside the inclusion is uniform. However, the lower bound can only be approached since no boundary data generate a nonzero uniform field outside the inclusion. The lower bound (for $f_1$) can be attained for the configuration where phase 2 is an inclusion.

There are plenty of inclusions inside which the field is uniform for some boundary conditions. We call such inclusions E$_\Om$-inclusions.
They include E-inclusions which were named in \cite{ljl07}.
An inclusion $E$ is called an E-inclusion if the field inside $E$ is uniform for any uniform loading at infinity. More precisely, E-inclusions are such that if $V$ is the solution to
 \beq
 \left\{
 \begin{array}{l}
 \Grad \cdot (\Gs_1 \chi(E) + \Gs_2 \chi(\RR^2 \setminus E)) \Grad V=0  \quad  \mbox{in } \RR^2, \\
 V(\Bx) - \Ba \cdot \Bx = O(|\Bx|^{-1}) \ \ \mbox{as } |\Bx| \to \infty,
 \end{array}
 \right.
 \eeq{In6}
then $-\Grad V$ is constant in $E$ for any direction $\Ba$.  If an E-inclusion $E$ is simply connected, then $E$ must be an ellipse (an ellipsoid in three dimensions). This was known as Eshelby's conjecture \cite{esh61} and resolved by Sendeckyj in two dimensions \cite{sen70} (see also \cite{KM06, liu07} for different proofs), and by Kang-Milton \cite{KM07} and Liu \cite{liu07} in three dimensions. There are E-inclusions with multiple components \cite{che74, liu07, KKM07}. There are also inclusions other than E-inclusions inside which the field is uniform. For example, if $\Om$ contains a connected component, say $E$, of an E-inclusion with multiple components, then $E$ is an  E$_\Om$-inclusion. More generally if $E$ is an E$_\Om$-inclusion and $\GY\subset\GO$ then the field in $E\cap\GY$ will be uniform when appropriate boundary conditions are imposed at the boundary of $\GY$.

We perform some numerical experiments to demonstrate how good the bounds are for inclusions. Special attention is paid to the variation of the bounds when certain parameters, such as conductivity, the volume fraction and the distance from the boundary, vary. We also look at the role of boundary data.

This paper is organized as follows. In the next section we derive the lower and upper bounds on the volume fraction. In section 3, we obtain conditions for these bounds to be attained, and then in section 4, we show that if the field is uniform in phase 1 then the upper bound is attained and if the field is uniform in phase 2 then the lower bound is attained. In Section 5, we obtain different sufficient conditions for the bounds to be obtained. Section 6 is devoted to the asymptotic analysis of the bounds when the volume fraction tends to zero. Numerical results are presented in section 7. In section 8 we show how to construct a wide variety of simply
connected E$_\Om$-inclusions, following the approach outlined in section 23.9 of \cite{milton}.

We emphasize that the method of this paper (the translation method) works for three dimensions as well. The results in three dimensions will be
presented in a forthcoming paper.

\section{Translation bounds in two dimensions}
\setcounter{equation}{0}
In this section we derive upper and lower bounds on $f_1$ (the volume fraction of the phase with higher conductivity) using pairs of Cauchy data. Each bound requires two pairs of Cauchy data. 
The derivation in this section is based on the translation method, and parallels the treatment given by Murat and Tartar \cite{tar79,tar85,mutar85} and Lurie
and Cherkaev \cite{lucherk82,lucherk84}.

\subsection{Lower bound}

Consider two potentials satisfying
 \beq
 \Div \Gs \Grad V_j =0 \quad \mbox{in } \Om, \quad j=1,2.
 \eeq{LB1}
Let
 \beq
 \Bj_j (\Bx) = - \Gs(\Bx) \Grad V_j(\Bx), \quad j=1,2.
 \eeq{LB2}
We want to use information about two pairs of Cauchy data $(V_1^0, q_1:=-\Bj_1 \cdot \Bn)$ and $(V_2^0, q_2:=-\Bj_2 \cdot \Bn)$ on $\p\Om$ to generate a lower bound on $f_1$.

Using the boundary data we can compute
 \beq
 \lang \Bj_i \rang := \frac{1}{|\Om|} \int_{\Om} \Bj_i = -\frac{1}{|\Om|} \int_{\p\Om} \Bx q_i , \quad i=1,2.
 \eeq{LB3}
We assume that $\lang \Bj_1 \rang$ and $\lang \Bj_2 \rang$ are linearly independent. Then, by taking linear combinations of the old potentials if necessary we may assume
 \beq
 \lang \Bj_1 \rang = \begin{bmatrix} 1 \\ 0 \end{bmatrix}, \quad \lang \Bj_2 \rang = \begin{bmatrix} 0 \\ 1 \end{bmatrix} .
 \eeq{LB4}

With
 \beq
 R_\bot = \begin{bmatrix} 0 & 1 \\ -1 & 0 \end{bmatrix},
 \eeq{LB5}
let us introduce a $4 \times 4$ matrix
 \beq
 L_c(\Bx):= \begin{bmatrix} \Gs^{-1}  & c R_\bot \\
 - c R_\bot & \Gs^{-1}  \end{bmatrix}
 \eeq{LB6}
where the constant $c$ is chosen so that $L_c(\Bx) \ge 0$ for all $\Bx$. Here we assume that $\Gs$ is an anisotropic conductivity (matrix). With the constants $k_1$, $k_2$, $k_3$, $k_4$, define a 4-dimensional vector $J$ by
 \beq
 J:= \begin{bmatrix} k_1 \Bj_1 + k_2 \Bj_2 \\ k_3 \Bj_1 + k_4 \Bj_2 \end{bmatrix} .
 \eeq{LB7}
We then consider
 \beq
 W_c := \frac{1}{|\Om|} \int_\Om J \cdot L_c (\Bx) J.
 \eeq{LB8}

Define a $2 \times 2$ matrix $A=(a_{ij})$, which we call the response (or measurement) matrix, by
 \beq
 a_{ij} := \frac{1}{|\Om|} \int_\Om \Bj_i \cdot \Gs^{-1} \Bj_j = \frac{1}{|\Om|} \int_{\p\Om} V_i^0 q_j, \quad i,j=1,2,
 \eeq{LB9}
and
 \beq
 b:= \frac{1}{2|\Om|} \int_{\Om} \Bj_1 \cdot R_\bot \Bj_2 - \Bj_2 \cdot R_\bot \Bj_1 = \frac{1}{|\Om|} \int_{\Om} \Bj_1 \cdot R_\bot \Bj_2 .
 \eeq{LB10}
Since
 \beq
 \int_{\Om} \Bj_i \cdot R_\bot \Bj_i =0, \quad i=1,2,
 \eeq{LB11}
one can see that
 \beq
 W_c = \begin{bmatrix} k_1 \\ k_2 \\ k_3 \\ k_4 \end{bmatrix} \cdot D_c \begin{bmatrix} k_1 \\ k_2 \\ k_3 \\ k_4 \end{bmatrix}
 \eeq{LB12}
where
 \beq
 D_c = \begin{bmatrix} a_{11} & a_{12} & 0 & cb \\ a_{12} & a_{22} & -cb & 0 \\ 0 & -cb & a_{11} & a_{12} \\
 cb & 0 & a_{12} & a_{22} \end{bmatrix} .
 \eeq{LB13}
We emphasize that $W_c$ can be computed from the boundary measurements. In fact, since $\Curl R_{\bot} \Bj_i =0$, there are potentials $\ph_i$ such that
 \beq
 R_{\bot} \Bj_i = \Grad \ph_i.
 \eeq{LB14}
Moreover, if $\Bt$ is the unit tangent vector field on $\p\Om$ in the positive orientation, then
 \beq
 \Bt \cdot \Grad \ph_i = R_\bot^T \Bt \cdot \Bj_i = - \Bj_i \cdot \Bn = q_i \quad \mbox{on } \p\Om
 \eeq{LB15}
(T for the transpose), and hence the boundary value of $\ph_i$ which we denote by $\ph_i^0$ is given by
 \beq
 \ph_i^0 (\Bx)= \int_{\Bx_0}^{\Bx} q_i
 \eeq{LB16}
where the integration is along $\p\Om$ in the positive orientation (counterclockwise). Hence
 \beq
 b = - \frac{1}{|\Om|} \int_{\p\Om} q_1 \ph_2^0 = \frac{1}{|\Om|} \int_{\p\Om} q_2 \ph_1^0.
 \eeq{LB17}

Since
 \begin{align}
 W_c & = \frac{1}{|\Om|} \int_{\Om} (k_1 \Bj_1 + k_2 \Bj_2) \Gs^{-1} (k_1 \Bj_1 + k_2 \Bj_2) + (k_3 \Bj_1 + k_4 \Bj_2) \Gs^{-1} (k_3 \Bj_1 + k_4 \Bj_2) \nonumber \\
 &\quad + 2c (k_1k_4 -k_2k_3) b, \label{LB18}
 \end{align}
we have the variational principle
 \beq
 W_c= \min_{\dis \Div \underline{\Bj_1} = \Div\underline{\Bj_2}=0 \atop \dis \underline{\Bj_1} \cdot \Bn= q_1 , \ \underj_2 \cdot \Bn= q_2} \frac{1}{|\Om|} \int_{\Om} \begin{bmatrix} k_1 \underline{\Bj_1} + k_2 \underline{\Bj_2} \\ k_3 \underline{\Bj_1} + k_4 \underline{\Bj_2} \end{bmatrix} \cdot L_c(\Bx) \begin{bmatrix} k_1 \underline{\Bj_1} + k_2 \underline{\Bj_2} \\ k_3 \underline{\Bj_1} + k_4 \underline{\Bj_2} \end{bmatrix} .
 \eeq{LB19}
One can easily see from the constraints that
 \beq
 \lang \underline{\Bj_i} \rang =\frac{1}{|\Om|} \int_{\p\Om} -\Bx q_i = \lang \Bj_i \rang .
 \eeq{LB20}
So if we replace the constraints by the weaker constraint that
 \beq
 \lang \underline{\Bj_i} \rang = \lang \Bj_i \rang , \quad i=1,2,
 \eeq{LB21}
then we get
 \beq
  W_c \ge \min_{\dis \underline{\Bj_1}, \underline{\Bj_2} \atop \dis \lang \underline{\Bj_i} \rang = \lang \Bj_i \rang} \left\lang \begin{bmatrix} k_1 \underline{\Bj_1} + k_2 \underline{\Bj_2} \\ k_3 \underline{\Bj_1} + k_4 \underline{\Bj_2} \end{bmatrix} \cdot L_c \begin{bmatrix} k_1 \underline{\Bj_1} + k_2 \underline{\Bj_2} \\ k_3 \underline{\Bj_1} + k_4 \underline{\Bj_2} \end{bmatrix} \right\rang .
 \eeq{LB22}
In order to find the minimum, we first observe that at the minimum
 \beq
 \int_{\Om} \begin{bmatrix} k_1 \psi_1 + k_2 \psi_2 \\ k_3 \psi_1 + k_4 \psi_2 \end{bmatrix} \cdot L_c (\Bx) \begin{bmatrix} k_1 \underline{\Bj_1} + k_2 \underline{\Bj_2} \\ k_3 \underline{\Bj_1} + k_4 \underline{\Bj_2} \end{bmatrix} =0
 \eeq{LB23}
for any (vector-valued) functions $\psi_1, \psi_2$ satisfying $\lang \psi_1 \rang = \lang \psi_2 \rang =0$, which implies
 \beq
 \dis L_c (\Bx) \begin{bmatrix} k_1 \underline{\Bj_1} + k_2 \underline{\Bj_2} \\ k_3 \underline{\Bj_1} + k_4 \underline{\Bj_2} \end{bmatrix} = \mu \ (\mbox{a constant vector}).
 \eeq{LB24}
We then have
 \beq
 \left\lang \begin{bmatrix} k_1 \Bj_1 + k_2 \Bj_2 \\ k_3 \Bj_1 + k_4 \Bj_2 \end{bmatrix} \right\rang = \left\lang \begin{bmatrix} k_1 \underline{\Bj_1} + k_2 \underline{\Bj_2} \\ k_3 \underline{\Bj_1} + k_4 \underline{\Bj_2} \end{bmatrix} \right\rang = \lang L_c^{-1} \rang \mu
 \eeq{LB25}
Thus the minimum is given by
 \begin{align}
 & \left\lang \begin{bmatrix} k_1 \underline{\Bj_1} + k_2 \underline{\Bj_2} \\ k_3 \underline{\Bj_1} + k_4 \underline{\Bj_2} \end{bmatrix} \cdot L_c \begin{bmatrix} k_1 \underline{\Bj_1} + k_2 \underline{\Bj_2} \\ k_3 \underline{\Bj_1} + k_4 \underline{\Bj_2} \end{bmatrix} \right\rang \nonumber \\
 & = \lang \mu \cdot L_c^{-1} \mu \rang = \left \lang \begin{bmatrix} k_1 \Bj_1 + k_2 \Bj_2 \\ k_3 \Bj_1 + k_4 \Bj_2 \end{bmatrix} \right\rang \cdot \lang L_c^{-1} \rang^{-1} \left\lang \begin{bmatrix} k_1 \Bj_1 + k_2 \Bj_2 \\ k_3 \Bj_1 + k_4 \Bj_2 \end{bmatrix} \right\rang, \label{LB26}
 \end{align}
which implies, thanks to (\ref{LB4}), that
 \beq
 W_c \ge \begin{bmatrix} k_1 \\ k_2 \\ k_3 \\ k_4 \end{bmatrix} \cdot \lang L_c^{-1} \rang^{-1} \begin{bmatrix} k_1 \\ k_2 \\ k_3 \\ k_4 \end{bmatrix} .
 \eeq{LB27}
Thus we have
 \beq
 D_c \ge \lang L_c^{-1} \rang^{-1} .
 \eeq{LB28}

Let us now assume that $\Gs$ is isotropic so that
 \beq
 L_c= \begin{bmatrix} \Gs^{-1} & 0 & 0 & c \\
 0 & \Gs^{-1} & -c & 0 \\
 0 & -c & \Gs^{-1} & 0 \\
 c & 0 & 0 & \Gs^{-1}
 \end{bmatrix} ,
 \eeq{LB29}
and
 \beq
 \lang L_c^{-1} \rang = \left\lang \frac{1}{(\Gs^{-2}-c^2)} \begin{bmatrix} \Gs^{-1} & 0 & 0 & -c \\
 0 & \Gs^{-1} & c & 0 \\
 0 & c & \Gs^{-1} & 0 \\
 -c & 0 & 0 & \Gs^{-1}
 \end{bmatrix} \right\rang .
 \eeq{LB30}
Since
 \beq
 \begin{bmatrix}
 Q^T & 0 \\
 0 & Q^T
 \end{bmatrix} \lang L_c^{-1} \rang^{-1} \begin{bmatrix}
 Q & 0 \\
 0 & Q
 \end{bmatrix} = \lang L_c^{-1} \rang^{-1}
 \eeq{LB31}
for any rotation $Q$, we  obtain from (\ref{LB28}) that
 \beq
 \begin{bmatrix}
 Q^T & 0 \\
 0 & Q^T
 \end{bmatrix} D_c \begin{bmatrix}
 Q & 0 \\
 0 & Q
 \end{bmatrix} \ge \lang L_c^{-1} \rang^{-1} .
 \eeq{LB32}
In particular, we may choose $Q$ so that
 \beq
 Q^T \begin{bmatrix} a_{11} & a_{12} \\
 a_{12} & a_{22} \end{bmatrix} Q = \begin{bmatrix} \la_1 & 0 \\
 0 & \la_2 \end{bmatrix}
 \eeq{LB33}
where $\la_1 \ge \la_2$ are eigenvalues of the response matrix $(a_{ij})$. Then by taking the inverse of both sides of (\ref{LB32}) we get
 \beq
 \frac{1}{(\la_1 \la_2 - c^2 b^2)} \begin{bmatrix} \la_2 & 0 & 0 & -cb \\ 0 & \la_1 & cb & 0 \\ 0 & cb & \la_2 & 0 \\
 -cb & 0 & 0 & \la_1 \end{bmatrix} \le \lang L_c^{-1} \rang .
 \eeq{LB34}
So we get the inequality
 \beq
 \frac{1}{(\la_1 \la_2 - c^2 b^2)} \Bv \cdot \begin{bmatrix} \la_2 & -cb \\ -cb & \la_1 \end{bmatrix} \Bv \le
 \left\lang \frac{1}{(\Gs^{-2} - c^2)} \Bv \cdot \begin{bmatrix} \Gs^{-1} & -c \\ -c & \Gs^{-1} \end{bmatrix} \Bv \right \rang
 \eeq{LB35}
for any vector $\Bv$.

Now suppose that the medium is 2-phase, with $\Gs_1 > \Gs_2$. In this case $L_c(\Bx) > 0$ as long as $c < \Gs_1^{-1}$. We take the limit as $c$ approaches $\Gs_1^{-1}$. Then
 \beq
 \frac{\Bv}{(\Gs_1^{-2} - c^2)} \cdot \begin{bmatrix} \Gs_1^{-1} & -c \\ -c & \Gs_1^{-1} \end{bmatrix} \Bv
 \eeq{LB36}
becomes infinite unless $\Bv$ is proportional to $\begin{bmatrix} 1 \\ 1 \end{bmatrix}$, and when $\Bv=\begin{bmatrix} 1 \\ 1 \end{bmatrix}$
 \beq
 \frac{\Bv}{(\Gs^{-2} - c^2)} \cdot \begin{bmatrix} \Gs^{-1} & -c \\ -c & \Gs^{-1} \end{bmatrix} \Bv
 = \frac{2(\Gs^{-1} -c)}{\Gs^{-2}-c^2}= \frac{2}{\Gs^{-1} + c}
 \eeq{LB37}
approaches $\Gs_1$ in phase 1 and $2/(\Gs_1^{-1}+\Gs_2^{-1})$ in phase 2. Hence the bound in (\ref{LB35}) reduces to
 \begin{align}
 \frac{\la_1 + \la_2 - 2b/\Gs_1}{\la_1 \la_2 - b^2/\Gs_1^2}
 & \le f_1 \Gs_1 + \frac{2f_2}{1/\Gs_1 + 1/\Gs_2} \\
 & = f_1 \Gs_1 + \frac{2f_2 \Gs_1 \Gs_2}{\Gs_1 + \Gs_2} \\
 & = f_1 \frac{\Gs_1 (\Gs_1 -\Gs_2)}{(\Gs_1 +\Gs_2)} + \frac{2 \Gs_1 \Gs_2}{\Gs_1 + \Gs_2},
\label{LB37a}
 \end{align}
which gives the desired lower bound on the volume fraction:
 \beq
 f_1 \ge \frac{(\Gs_1 + \Gs_2)}{\Gs_1 (\Gs_1 - \Gs_2)} \left[ \frac{\la_1 + \la_2 - 2b/\Gs_1}{\la_1 \la_2 - b^2/\Gs_1^2} - \frac{2 \Gs_1 \Gs_2}{\Gs_1 + \Gs_2} \right],
 \eeq{LB38}
or
 \beq
  f_1 \ge \frac{(\Gs_1 + \Gs_2)}{\Gs_1 (\Gs_1 - \Gs_2)} \left[ \frac{\Tr A - 2b/\Gs_1}{\det A - b^2/\Gs_1^2} -
 \frac{2 \Gs_1 \Gs_2}{\Gs_1 + \Gs_2} \right],
 \eeq{LB39}
where the matrix $A$ is defined by (\ref{LB9}). We emphasize that the righthand side of (\ref{LB39}) can be computed by the boundary measurements. In fact, $A$ is computed by using (\ref{LB9}) and $b$ using (\ref{LB17}) under the condition (\ref{LB4}).

In general, if Neumann data $q_1$ and $q_2$ do not satisfy \eq{LB4}, then let
 \beq
 P_N:= \begin{bmatrix} \dis \frac{-1}{|\Om|} \int_{\p\Om} q_1 \Bx & \ \ \dis \frac{-1}{|\Om|} \int_{\p\Om} q_2 \Bx \end{bmatrix}^{-1} .
 \eeq{LB40}
Then $\tilde{\Bj}_1$ and $\tilde{\Bj}_2$ defined by
 \beq \tilde{\Bj}_i=\sum_{m=1}^2\left[ P_N\right]_{im} \Bj_m,\quad i=1,2
 \eeq{LB41}
satisfy \eq{LB4}. Since
 \beq
 \left[ \frac{1}{|\Om|} \int_{\Om} \tilde{\Bj_i} \cdot \Gs^{-1} \tilde{\Bj_j} \right]_{i,j=1,2} = P_N A P_N^T
 \eeq{LB42}
and
 \beq
 \frac{1}{|\Om|} \int_{\Om} \tilde{\Bj}_1 \cdot R_\bot \tilde{\Bj}_2 = \frac{\det P_N}{|\Om|} \int_{\Om} \Bj_1 \cdot R_\bot \Bj_2,
 \eeq{LB43}
we obtain the following theorem from \eq{LB39}.

\begin{theorem}\label{thm:LB1}
Let $P_N$ be given by \eq{LB40} and
 \beq
 b_N := \frac{1}{|\Om|} \int_{\Om} \Bj_1 \cdot R_\bot \Bj_2 = \frac{1}{|\Om|} \int_{\p\Om} q_1(\Bx) (\int_{\Bx_0}^{\Bx} q_2).
 \eeq{LB44}
Then,
 \beq
  f_1 \ge \frac{(\Gs_1 + \Gs_2)}{\Gs_1 (\Gs_1 - \Gs_2)} \left[ \frac{\Tr (P_N A P_N^T) - 2 (\det P_N)b_N /\Gs_1}{(\det P_N)^2 (\det A - b_N^2/\Gs_1^2)} -
 \frac{2 \Gs_1 \Gs_2}{\Gs_1 + \Gs_2} \right].
 \eeq{LB45}
\end{theorem}

\subsection{Upper bound}

We now derive the upper bound on $f_1$.

Let us introduce a $4 \times 4$ matrix
 \beq
 L'_c (\Bx):= \begin{bmatrix} \Gs  & c R_\bot \\
 - c R_\bot & \Gs  \end{bmatrix}
 \eeq{UB1}
where the constant $c$ is chosen so that $L'_c(\Bx) \ge 0$ for all $\Bx$. With the constants $k_1$, $k_2$, $k_3$, $k_4$ and
 \beq
 \Be_j (\Bx) = - \Grad V_j(\Bx), \quad j=1,2,
 \eeq{UB2-1}
define a 4-dimensional vector $E$ by
 \beq
 E:= \begin{bmatrix} k_1 \Be_1 + k_2 \Be_2 \\ k_3 \Be_1 + k_4 \Be_2 \end{bmatrix} .
 \eeq{UB2}
We then consider
 \beq
 W'_c:= \lang E \cdot L'_c E \rang.
 \eeq{UB3}
The minimization problem in this case is
 \beq
  W'_c \ge \min_{\dis \underline{\Be_1}, \underline{\Be_2} \atop \dis \lang \underline{\Be_i} \rang = \lang \Be_i \rang} \left\lang \begin{bmatrix} k_1 \underline{\Be_1} + k_2 \underline{\Be_2} \\ k_3 \underline{\Be_1} + k_4 \underline{\Be_2} \end{bmatrix} \cdot L'_c \begin{bmatrix} k_1 \underline{\Be_1} + k_2 \underline{\Be_2} \\ k_3 \underline{\Be_1} + k_4 \underline{\Be_2} \end{bmatrix} \right\rang .
 \eeq{UB3-1}
As for \eq{LB24}, one can show that at the minimum of the right hand side of \eq{UB3-1}
 \beq
 \dis L'_c (\Bx) \begin{bmatrix} k_1 \underline{\Be_1} + k_2 \underline{\Be_2} \\ k_3 \underline{\Be_1} + k_4 \underline{\Be_2} \end{bmatrix} = \Gm \ (\mbox{a constant vector})
 \eeq{UB3-2}
and the minimum is given by
 \begin{align}
 \left\lang \begin{bmatrix} k_1 \underline{\Be_1} + k_2 \underline{\Be_2} \\ k_3 \underline{\Be_1} + k_4 \underline{\Be_2} \end{bmatrix} \cdot L'_c \begin{bmatrix} k_1 \underline{\Be_1} + k_2 \underline{\Be_2} \\ k_3 \underline{\Be_1} + k_4 \underline{\Be_2} \end{bmatrix} \right\rang
 = \left\lang \begin{bmatrix} k_1 \Be_1 + k_2 \Be_2 \\ k_3 \Be_1 + k_4 \Be_2 \end{bmatrix} \right\rang \cdot \lang (L'_c)^{-1} \rang^{-1} \left\lang \begin{bmatrix} k_1 \Be_1 + k_2 \Be_2 \\ k_3 \Be_1 + k_4 \Be_2 \end{bmatrix} \right\rang. \label{UB3-3}
 \end{align}

Proceeding in the exactly same way as in the previous subsection (with $c$ approaching to $\Gs_2$), we can derive `dual bounds':
 \beq
 \frac{\Tr A - 2b' \Gs_2}{\det A - b'^2 \Gs_2^2} \le \frac{f_2}{\Gs_2} + \frac{2f_1}{\Gs_1+\Gs_2}
 \eeq{UB4}
where
 \beq
 b' := \lang \Be_1 \cdot R_\bot \Be_2 \rang
 \eeq{UB5}
and
 \beq
 A = \begin{bmatrix} a_{11} & a_{12} \\
 a_{12} & a_{22} \end{bmatrix}
 \eeq{UB6}
in which
 \beq
 a_{ij} := \lang \Be_i \cdot \Gs \Be_j \rang = \lang \Be_i \cdot \Bj_j \rang,
 \eeq{UB7}
and linear combination of potentials have been chosen so that
 \beq
 \lang \Be_1 \rang = \frac{1}{|\Om|} \int_{\p\Om} -V_1^0 \Bn = \begin{bmatrix} 1 \\ 0 \end{bmatrix} , \quad
 \lang \Be_2 \rang = \frac{1}{|\Om|} \int_{\p\Om} -V_2^0 \Bn = \begin{bmatrix} 0 \\ 1 \end{bmatrix}.
 \eeq{UB8}
Apart from this constraint, $\Be_1$ and $\Be_2$ are any fields solving
 \beq
 \Div \Gs \Grad V_j =0 \quad \mbox{in } \Om, \quad \Be_j = - \Grad V_j.
 \eeq{UB9}

One can obtain from (\ref{UB4}) the upper bound on $f_1$:
 \beq
 f_1 \le \frac{\Gs_2 (\Gs_1+\Gs_2)}{(\Gs_1-\Gs_2)} \left[ \frac{1}{\Gs_2} - \frac{\Tr A - 2b' \Gs_2}{\det A - b'^2 \Gs_2^2} \right].
 \eeq{UB10}
We emphasize that $A$ and $b'$ can be computed from the boundary measurements:
 \beq
 a_{ij}= \frac{1}{|\Om|} \int_{\p\Om} V_i^0 q_j,
 \eeq{UB11}
and
 \beq
 b' = \frac{1}{|\Om|} \int_{\p\Om} V_1^0 \Bn \cdot R_\bot \Be_2 = - \frac{1}{|\Om|} \int_{\p\Om} V_1^0 \Bt \cdot \Be_2 = \frac{1}{|\Om|} \int_{\p\Om} V_1^0 \frac{\p V_2^0}{\p \Bt}.
 \eeq{UB12}

More generally, if $V_1^0$ and $V_2^0$ do not satisfy \eq{UB8}, then we have the following theorem in the same way as before.
\begin{theorem}\label{thm:UB1}
Let
 \beq
 P_D:= \begin{bmatrix} \dis \frac{-1}{|\Om|} \int_{\p\Om} V_1^0 \Bn & \ \ \dis \frac{-1}{|\Om|} \int_{\p\Om} V_2^0 \Bn \end{bmatrix}^{-1}
 \eeq{UB20}
and
 \beq
 b_D := \frac{1}{|\Om|} \int_{\p\Om} V_1^0 \frac{\p V_2^0}{\p \Bt}.
 \eeq{UB21}
Then
 \beq
 f_1 \le \frac{\Gs_2 (\Gs_1+\Gs_2)}{(\Gs_1-\Gs_2)} \left[ \frac{1}{\Gs_2} - \frac{\Tr (P_D A P_D^T) - 2 (\det P_D) b_D \Gs_2}{(\det P_D)^2 (\det A - b_D^2 \Gs_2^2)} \right].
 \eeq{UB19}
\end{theorem}

\subsection{Special boundary data}

In the special case where the Neumann data are given by
 \beq
 q_1 = - \Bn \cdot \Bj_1 = - \Bn \cdot \begin{bmatrix} 1 \\ 0 \end{bmatrix}, \quad q_2 = - \Bn \cdot \Bj_2 = - \Bn \cdot \begin{bmatrix} 0 \\ 1 \end{bmatrix},
 \eeq{HS1}
we have
 \beq
 b=1 \quad\mbox{and}\quad A=\BGs_N^{-1}
 \eeq{HS2}
where $\BGs_N$ is the Neumann tensor which is defined via the relation
 \beq
 \lang \Be \rang = \BGs_N^{-1} \lang \Bj \rang,
 \eeq{HS3}
when the Neumann data is given by $q=-\Bn\cdot\Bv$ for some constant vector $\Bv$.
In fact, we have from \eq{LB14} and (\ref{LB17})
 \beq
 b= \frac{1}{|\Om|} \int_{\p\Om} (\Bj_1 \cdot \Bn) \ph_2^0 = \frac{1}{|\Om|} \Bj_1 \cdot \int_{\p\Om} \Bn \ph_2^0 =- \begin{bmatrix} 1 \\ 0 \end{bmatrix} \cdot \lang \Grad \ph_2 \rang = \begin{bmatrix} 1 \\ 0 \end{bmatrix} \cdot R_\bot \begin{bmatrix} 0 \\ 1 \end{bmatrix} =1,
 \eeq{HS4}
and from (\ref{LB9})
 \beq
 a_{ij} = \frac{1}{|\Om|} \int_{\p\Om} V_i^0 q_j = \frac{1}{|\Om|} \Bj_j^0 \cdot \int_{\p\Om} V_i^0 \Bn = \lang \Bj_i \rang \cdot \lang \Be_j \rang = \lang \Bj_i \rang \cdot \BGs_N^{-1} \lang \Bj_j \rang .
 \eeq{HS5}
So, the bound (\ref{LB39}) reduces to the bound
 \beq
  f_1 \ge \frac{(\Gs_1 + \Gs_2)}{\Gs_1 (\Gs_1 - \Gs_2)} \left[ \frac{\Tr \BGs_N^{-1} - 2/\Gs_1}{\det \BGs_N^{-1} - 1/\Gs_1^2} -
 \frac{2 \Gs_1 \Gs_2}{\Gs_1 + \Gs_2} \right].
 \eeq{HS6}
of Milton \cite{milt11}.

If the Dirichlet data take the special affine form
 \beq
 V_1^0 = - \begin{bmatrix} 1 \\ 0 \end{bmatrix} \cdot \Bx, \quad V_2^0 = - \begin{bmatrix} 0 \\ 1 \end{bmatrix} \cdot \Bx,
 \eeq{HS7}
one can prove in the same way that
 \beq
 b'=1 \quad\mbox{and}\quad A=\BGs_D
 \eeq{HS8}
where $\BGs_D$ is the Dirichlet tensor which is defined via the relation
 \beq
 \BGs_D \lang \Be \rang = \lang \Bj \rang
 \eeq{HS9}
when the Dirichlet data $V^0$ is given by $-\Bv \cdot \Bx$ for some constant vector $\Bv$. Thus the bound (\ref{UB10}) reduces to the other bound
 \beq
 f_1 \le \frac{\Gs_2 (\Gs_1+\Gs_2)}{(\Gs_1-\Gs_2)} \left[ \frac{1}{\Gs_2} - \frac{\Tr \BGs_D - 2 \Gs_2}{\det \BGs_D - \Gs_2^2} \right].
 \eeq{HS10}
of Milton \cite{milt11}.

\section{Attainability conditions of the bounds}

In this section we derive conditions on the fields for the bounds in \eq{LB39} and \eq{UB10} to be attained. We will show in the next section that the bounds are actually attained by certain inclusions.

The derivation of the lower bound on $f_1$, and in particular \eq{LB24} and \eq{LB25}, suggests that if there is no column vector $\BK=(k_1,k_2,k_3,k_4)^T$, with say
$|\BK|^2=k_1^2+k_2^2+k_3^2+k_4^2=1$, such that
\beq L_{\Gs_1^{-1}} \begin{bmatrix} k_1 {\Bj_1} + k_2 {\Bj_2} \\ k_3 {\Bj_1} + k_4 {\Bj_2} \end{bmatrix}  = \lang (L_{\Gs_1^{-1}})^{-1} \rang^{-1} \left\lang \begin{bmatrix} k_1 {\Bj_1} + k_2 {\Bj_2} \\ k_3 {\Bj_1} + k_4 {\Bj_2} \end{bmatrix} \right\rang ,
\eeq{ACB1}
then the lower bound will not be attained. Here, $\lang (L_{\Gs_1^{-1}})^{-1} \rang^{-1}$ is understood as the limit of $\lang L_{c}^{-1} \rang^{-1}$
as $c$ tends to $\Gs_1^{-1}$. To prove this, fix $c_0 < \Gs_1^{-1}$ and let
 \beq
 \BF_c (\Bx): =  L_{c}(\Bx) \begin{bmatrix} k_1 {\Bj_1} + k_2 {\Bj_2} \\ k_3 {\Bj_1} + k_4 {\Bj_2} \end{bmatrix} - \lang L_{c}^{-1} \rang^{-1} \left\lang \begin{bmatrix} k_1 {\Bj_1} + k_2 {\Bj_2} \\ k_3 {\Bj_1} + k_4 {\Bj_2} \end{bmatrix} \right\rang.
 \eeq{ACB2}
for $c$ such that $c_0 \le c < \Gs_1^{-1}$. Then, we have
 \begin{align}
 & \lang \BF_c \cdot L_{c_0}^{-1} \BF_c \rang \le \lang \BF_c \cdot L_{c}^{-1} \BF_c \rang \nonumber \\
 & =  \left\lang \begin{bmatrix} k_1 {\Bj_1} + k_2 {\Bj_2} \\ k_3 {\Bj_1} + k_4 {\Bj_2} \end{bmatrix} \cdot L_{c} \begin{bmatrix} k_1 {\Bj_1} + k_2 {\Bj_2} \\ k_3 {\Bj_1} + k_4 {\Bj_2} \end{bmatrix} \right\rang - \left\lang \begin{bmatrix} k_1 {\Bj_1} + k_2 {\Bj_2} \\ k_3 {\Bj_1} + k_4 {\Bj_2} \end{bmatrix} \right\rang \cdot \lang L_{c}^{-1} \rang^{-1} \left\lang \begin{bmatrix} k_1 {\Bj_1} + k_2 {\Bj_2} \\ k_3 {\Bj_1} + k_4 {\Bj_2} \end{bmatrix} \right\rang \nonumber \\
&=\BK\cdot D_{c}\BK-\BK\cdot\lang L_{c}^{-1}\rang^{-1}\BK \label{ACB3}
\end{align}
Letting $c\to \Gs_1^{-1}$ we see that if $\BF_{\Gs_1^{-1}}$ is non-zero (in the $L^2$ norm) for all $\BK$ with $|\BK|=1$ then the right hand side of \eq{ACB3} is non-zero in this limit, or equivalently
$D_{\Gs_1^{-1}} > \Ga\lang (L_{\Gs_1^{-1}})^{-1} \rang^{-1}$ for some $\Ga>1$. It follows that equality is not achieved in \eq{LB35} and hence in \eq{LB39}, {\it i.e.}, the lower bound on the volume fraction is not
attained.

Conversely, suppose we have equality in \eq{ACB1} for some $\BK\ne 0$. Then,
\beq \BK\cdot D_{\Gs_1^{-1}}\BK=\BK\cdot\lang (L_{\Gs_1^{-1}})^{-1}\rang^{-1}\BK,
\eeq{ACB4}
and as $D_{\Gs_1^{-1}} \ge \lang (L_{\Gs_1^{-1}})^{-1} \rang^{-1}$ it follows that
$D_{\Gs_1^{-1}}-\lang (L_{\Gs_1^{-1}})^{-1} \rang^{-1}$ must have zero determinant. A simple calculation shows that
\beq
 \lang (L_{\Gs_1^{-1}})^{-1} \rang^{-1} = \frac{1}{g} \begin{bmatrix} I & R_\bot \\ -R_\bot & I \end{bmatrix}
 \eeq{ACB5}
where
 \beq
 g:=   f_1 \frac{\Gs_1(\Gs_1 - \Gs_2)}{\Gs_1 + \Gs_2}+
 \frac{2 \Gs_1\Gs_2}{\Gs_1 + \Gs_2}.
 \eeq{ACB6}
Hence the matrix
\beq
\begin{bmatrix} Q^T & 0 \\ 0 & Q^T \end{bmatrix}[D_{\Gs_1^{-1}}-\lang (L_{\Gs_1^{-1}})^{-1} \rang^{-1}]\begin{bmatrix} Q & 0 \\ 0 & Q \end{bmatrix}
= \begin{bmatrix} \Gl_1-1/g & 0 & 0 & b/\Gs_1-1/g \\ 0 & \Gl_2-1/g & -b/\Gs_1+1/g & 0 \\ 0 & -b/\Gs_1+1/g & \Gl_1-1/g & 0 \\
 b/\Gs_1-1/g & 0 & 0 & \Gl_2-1/g \end{bmatrix}
\eeq{ACB7}
must have zero determinant, which implies
\beq \Gl_1\Gl_2-(\Gl_1+\Gl_2)/g-b^2/\Gs_1^2+2b/(g\Gs_1)=0.
\eeq{ACB8}
Thus equality holds in \eq{LB37a} and the lower bound on $f_1$ is attained.

In summary, the attainability condition is that for some $k_1$, $k_2$, $k_3$ and $k_4$,
 \beq
 L_{\Gs_1^{-1}} \BJ  = \lang (L_{\Gs_1^{-1}})^{-1} \rang^{-1} \lang \BJ \rang
 \eeq{AC8}
where
 \beq
 \BJ= \begin{bmatrix} k_1 {\Bj_1} + k_2 {\Bj_2} \\ k_3 {\Bj_1} + k_4 {\Bj_2} \end{bmatrix} .
 \eeq{AC9}
From \eq{ACB5} a vector $\BU$ in the range of $\lang (L_{\Gs_1^{-1}})^{-1} \rang^{-1}$ takes the form
 \beq
 \BU= \begin{bmatrix} a_1 \\ a_2 \\ -a_2 \\ a_1\end{bmatrix}
 \eeq{AC12}
for some $a_1$ and $a_2$.

We have the following theorem.
\begin{theorem}
The attainability condition \eq{AC8} for the lower bound holds if and only if
 \beq
 L_{\Gs_1^{-1}} \BJ = \BU
 \eeq{AC13}
for some $\BU$ of the form \eq{AC12}.
\end{theorem}

\proof The `only if' part is trivial. Suppose that \eq{AC13} holds. We write $L=L_{\Gs_1^{-1}}$ for the ease of notation. One can see from the definition \eq{LB6} of $L$ that $L_1$ (=$L$ on phase 1) and $L_2$ (=$L$ on phase 2) can be simultaneously diagonalizable. Thus in that basis \eq{AC13} reads as
 \beq
 \left[ \Gl_1^{(j)} \chi_1(\Bx) + \Gl_2^{(j)} \chi_2(\Bx) \right] J^{(j)}(\Bx) = U^{(j)}, \quad j=1,2,3,4.
 \eeq{AC14}
Here $\Gl_1^{(j)}$ and $\Gl_2^{(j)}$ are eigenvalues of $L_1$ and $L_2$, respectively,
and $J^{(j)}(\Bx)$ and $U^{(j)}$ are $j$-th components of $\BJ$ and $\BU$ in new basis.

Since $L_1$ has rank 2, two of eigenvalues $\Gl_1^{(j)}$ are zero, say $\Gl_1^{(3)}$ and $\Gl_1^{(4)}$, and hence $\chi_1 J^{(3)}(\Bx)$ and $\chi_2 J^{(4)}(\Bx)$ may depend on $\Bx$. However, $J^{(j)}(\Bx)$ for $j=1,2$ is piecewise constant, and by \eq{AC14},
 \beq
 J^{(j)}(\Bx) = \left\{
 \begin{array}{l}
 U^{(j)}/\Gl_1^{(j)} \quad \mbox{in phase 1} \\
 U^{(j)}/\Gl_2^{(j)} \quad \mbox{in phase 2} .
 \end{array}
 \right.
 \eeq{AC16}
Thus we have
 \beq
 \lang J^{(j)} \rang = \left[ f_1 /\Gl_1^{(j)} + f_2 /\Gl_2^{(j)} \right] U^{(j)} = \lang L^{-1} \rang_{jj} U^{(j)}, \quad j=1,2.
 \eeq{AC17}
Here $\lang L^{-1} \rang_{jj}$ is the $(j,j)$-entry of the diagonal matrix $\lang L^{-1} \rang$. So,
 \beq
 U^{(j)}= (\lang L^{-1} \rang^{-1})_{jj} \lang J^{(j)} \rang, \quad j=1,2.
 \eeq{AC18}
If $j=1,2$, then $(\lang L^{-1} \rang^{-1})_{jj}=0$ and $\BU$, which belongs to the range of $\lang L^{-1} \rang^{-1}$ satisfies $U^{(3)}=U^{(4)}=0$, and hence \eq{AC18} holds for all $j$. Therefore
 \beq
 \BU= \lang L^{-1} \rang^{-1} \lang \BJ \rang,
 \eeq{AC19}
and hence \eq{AC8} holds. \qed

Similarly one can show that the attainability condition for the upper bound is that for some $k_1$, $k_2$, $k_3$ and $k_4$
 \beq
 L'_{\Gs_2} \BE  = \lang (L'_{\Gs_2})^{-1} \rang^{-1} \lang \BE \rang
 \eeq{AC20}
where
 \beq
 \BE = \begin{bmatrix} k_1 {\Be_1}+ k_2 \Be_2 \\ k_3 \Be_1 + k_4 {\Be_2} \end{bmatrix},
 \eeq{AC21}
and it is equivalent to
 \beq
 L'_{\Gs_2} \BE = \BU
 \eeq{AC22}
for some $\BU$ of the form \eq{AC12}. We emphasize that the attainability conditions \eq{AC13} and \eq{AC22} are precisely analogous to those found by Grabovsky \cite{grab} for
composites.

\section{Attainability and uniformity}

We now investigate the attainability condition more closely. \eq{AC22} says that the field $\BE$ is uniform in phase 1. This condition alone guarantees that the upper bound is attained. In fact, we show in this section that even more is true: if the field is uniform in phase 1 for a single boundary data $V^0=V^0_1$ then there is a $V^0_2$ such that the upper bound is attained.

\begin{theorem}\label{thm:AU1}
Suppose that phase 1 and 2 have finitely many connected (possibly multiply connected) components and the interfaces are Lipschitz continuous. Let $V$ be the solution to
\beq
 \left\{
 \begin{array}{l}
 \Grad \cdot \Gs \Grad V=0  \quad  \mbox{in } \Om, \\
 V=V^0 \quad \mbox{on } \p\Om.
 \end{array}
 \right.
 \eeq{AU1}
If $-\Grad V$ is constant (the field is uniform) in phase 1 for some boundary data $V^0=V^0_1 \neq 0$, then there is a $V^0_2$ such the upper bound is attained.
\end{theorem}

\proof Phase 1 can be broken into connected components $\Psi_1^{(\alpha)}$, $\alpha= 1, 2, \ldots, m$, and phase 2 can be broken into connected components $\Psi_2^{(\Gb)}$, $\Gb= 1, 2, \ldots, n$.
If $\Psi_2^{(\Gb)}$ has a boundary in common with $\Psi_1^{(\alpha)}$, we denote the common boundary by $\Gamma^{\Ga\Gb}$.

Let $V_\Gb(\Bx)$ denote the potential $V(\Bx)$ inside $\Psi_2^{(\Gb)}$. If $-\Grad V= \begin{bmatrix} e_1 \\ e_2 \end{bmatrix}$ in phase 1 for some constants $e_1$ and $e_2$, then
 \beq
 V(\Bx) = -e_1 x - e_2 y + c_\Ga
 \eeq{AU2}
for some constant $c_\Ga$ inside $\Psi_1^{(\Ga)}$ (where $c_\Ga=c_\Gg$ if  $\Psi_1^{(\Ga)}$ touches $\Psi_1^{(\Gg)}$ at a common point),
and the continuity of the potential on $\Gamma^{\Ga\Gb}$ implies
 \beq
 V_\Gb (\Bx) = -e_1 x - e_2 y + c_\Ga \quad \mbox{on } \Gamma^{\Ga\Gb}.
 \eeq{AU3}
Since $\Grad \cdot \Bj(\Bx)=0$ in $\Om$, there is a continuous potential $W(\Bx)$ such that
 \beq
 \Bj (\Bx) = - \Gs_2 R_\bot \Grad W(\Bx) \quad \mbox{in } \Om.
 \eeq{AU4}
In phase 1, inside $\Psi_1^{(\Ga)}$, we have
 \beq
 \Grad W(\Bx) = \frac{\Gs_1}{\Gs_2} \begin{bmatrix} e_2 \\ -e_1 \end{bmatrix},
 \eeq{AU5}
and hence
 \beq
 W(\Bx) = \frac{\Gs_1}{\Gs_2} (e_2 x - e_1 y) + d_\Ga
 \eeq{AU6}
for some constant $d_\Ga$ (where, by continuity of the potential $W$, $d_\Ga=d_\Gg$ if  $\Psi_1^{(\Ga)}$ touches $\Psi_1^{(\Gg)}$ at a common point)

Let $W_\Gb(\Bx)$ denote the potential $W(\Bx)$ inside $\Psi_2^{(\Gb)}$. Since  $W(\Bx)$ is continuous,
 \beq
 W_\Gb (\Bx) = \frac{\Gs_1}{\Gs_2} (e_2 x - e_1 y) + d_\Ga \quad\mbox{on } \Gamma^{\Ga\Gb}.
 \eeq{AU8}
Note that inside $\Psi_2^{(\Gb)}$,
 \beq
 \Grad V(\Bx)= -\frac{1}{\Gs_2} \Bj(\Bx) = R_\bot \Grad W(\Bx),
 \eeq{AU9}
{\it i.e.}, $V_{\Gb, x}=W_{\Gb,y}$ and $V_{\Gb, y}=-W_{\Gb,x}$, which are the Cauchy-Riemann equations. Thus $V_\Gb+i W_\Gb$ is an analytic function of $z=x+iy$.

Now consider
 \begin{align}
 V'_\Gb(\Bx) &:= -W_\Gb(\Bx) + (\Gs_1/\Gs_2+1)(e_2 x - e_1  y) \label{AU10} \\
 W'_\Gb(\Bx) &:= V_\Gb(\Bx) + (\Gs_1/\Gs_2+1)(e_1 x + e_2  y). \label{AU11}
 \end{align}
Clearly
 \beq
 V'_\Gb + i W'_\Gb = i (V_\Gb + i W_\Gb) + (\Gs_1/\Gs_2+1)(e_2+ie_1)(x+iy)
 \eeq{AU12}
is an analytic function of $z$. On $\GG^{\Ga\Gb}$, we have
 \begin{align}
 V'_\Gb &= - (\Gs_1/\Gs_2) (e_2 x- e_1 y) + d_\Ga + (\Gs_1/\Gs_2+1)(e_2 x - e_1  y)  = e_2 x - e_1  y + d_\Ga, \label{AU13} \\
 W'_\Gb &= - e_1 x- e_2 y +c_\Ga + (\Gs_1/\Gs_2+1)(e_1 x + e_2  y)= (\Gs_1/\Gs_2) (e_1 x + e_2  y) + c_\Ga. \label{AU14}
 \end{align}
So the conductivity equation $\Grad \cdot \Gs \Grad V=0$ is satisfied with potentials $V'$ and $W'$ defined by $V'=V'_\Gb$, $W'=W'_\Gb$ in $\Psi_2^{(\Gb)}$, and
 \beq
 V'(\Bx)= e_2 x - e_1  y + d_\Ga, \quad W'(\Bx)= (\Gs_1/\Gs_2) (e_1 x + e_2  y) + c_\Ga
 \eeq{AU15}
in $\Psi_2^{(\Ga)}$. Note that
 \beq
 -\Grad V'= \begin{bmatrix} -e_2 \\ e_1 \end{bmatrix}.
 \eeq{AU16}

We then have, in $\Psi_2^{(\Gb)}$
 \begin{align}
 L'_{\Gs_2} \BE & = \begin{bmatrix} \Gs_2 I & \Gs_2 R_\bot \\ - \Gs_2 R_\bot & \Gs_2 I \end{bmatrix} \begin{bmatrix} -\Grad V \\ -\Grad V' \end{bmatrix} = \begin{bmatrix} \Gs_2 I & \Gs_2 R_\bot \\ - \Gs_2 R_\bot & \Gs_2 I \end{bmatrix} \begin{bmatrix} -\Grad V_\Gb \\ \Grad W_\Gb - (\Gs_1/\Gs_2 +1) \begin{bmatrix} e_2 \\ - e_1 \end{bmatrix} \end{bmatrix} \nonumber \\
 & = \begin{bmatrix} - \Gs_2 (\Grad V_\Gb - R_\bot \Grad W_\Gb) + (\Gs_1 + \Gs_2) \begin{bmatrix} e_1 \\ e_2 \end{bmatrix} \\ \Gs_2 (R_\bot \Grad V_\Gb + \Grad W_\Gb) + (\Gs_1 + \Gs_2) \begin{bmatrix} -e_2 \\ e_1 \end{bmatrix} \end{bmatrix} = (\Gs_1 + \Gs_2) \begin{bmatrix} e_1 \\ e_2 \\ -e_2 \\ e_1 \end{bmatrix}, \label{AU17}
 \end{align}
and in phase 1
 \begin{align}
 L'_{\Gs_2} \BE & = \begin{bmatrix} \Gs_1 I & \Gs_2 R_\bot \\ - \Gs_2 R_\bot & \Gs_1 I \end{bmatrix} \begin{bmatrix} -\Grad V \\ -\Grad V' \end{bmatrix} = \begin{bmatrix} \Gs_1 I & \Gs_2 R_\bot \\ - \Gs_2 R_\bot & \Gs_1 I \end{bmatrix} \begin{bmatrix} \begin{bmatrix} e_1 \\ e_2 \end{bmatrix} \\ \begin{bmatrix} -e_2 \\ e_1 \end{bmatrix} \end{bmatrix} \nonumber \\
 & = (\Gs_1 + \Gs_2) \begin{bmatrix} e_1 \\ e_2 \\ -e_2 \\ e_1 \end{bmatrix}.  \label{AU18}
 \end{align}
Thus $L'_{\Gs_2} \BE=\BU$ where $\BU$ is of the form \eq{AC12}. Hence the upper bound is attained when we take boundary data $V^0_1=V^0$ and $V^0_2=V'^0$ \qed

Observe that the Dirichlet condition in \eq{AU1} may be replaced with the Neumann condition. One can prove in the exactly same way that the lower bound is attained if the field is uniform in phase 2.

\section{Attainability and analyticity}
We have seen that uniformity and independence of the fields $\Be_1=-\Grad V_1$ and $\Be_2=-\Grad V_2$ in phase 1 is necessary and sufficient
to ensure that the upper bound is attained. Now we will see there is a condition on the potentials $V_1$ and $V_2$
in phase 2  which is also necessary and sufficient to ensure that the upper bound is attained. We assume that phase 2
is connected and completely surrounds each inclusion of phase 1.

First suppose that the upper bound is attained.
Then, given a constant $k$, there exist potentials $V$ and $V'$, which are linear combinations of the
potentials $V_1$ and $V_2$,
such that in phase 1 $-\Grad V= \begin{bmatrix} k \\ 0 \end{bmatrix}$ and
$-\Grad V'= \begin{bmatrix} 0 \\ k \end{bmatrix}$. Thus the analysis of the previous section holds
with $e_1=k$ and $e_2=0$. In particular, we may choose $k=1/(\Gs_1/\Gs_2+1)$ and, since in phase two $V+iW$
is an analytic function of $z$, it follows from \eq{AU10} that $V-iV'+x$ is an analytic function of $z=x+iy$ in phase 2.

Conversely suppose there exist potentials $V$ and $V'$, which are linear combinations of the
potentials $V_1$ and $V_2$, such that $V-iV'+x$ is an analytic function of $z$ in phase 2. Then the
harmonic conjugate to $V$ in phase 2 is $-V'-y$ and the harmonic conjugate to $V'$ in phase 2
is $V+x$. Since by \eq{AU9} these
harmonic conjugates can be identified with the potentials $W$ and $W'$, we have in phase 2
\beq W=-V'-y,\quad W'=V+x,
\eeq{AE12}
and in particular these identities hold on the boundary of an inclusion of phase 1. By \eq{AU4} inside that inclusion
$V+i(\Gs_2/\Gs_1)W$ and $V'+i(\Gs_2/\Gs_1)W'$ are analytic functions of $z$. Therefore
$$
V'+i(\Gs_2/\Gs_1)W'-i(\Gs_1/\Gs_2)(V+i(\Gs_2/\Gs_1)W)-i(x+iy)
$$
is also an analytic function of $z$ inside the inclusion and from \eq{AE12} takes the value
$$
i(\Gs_2/\Gs_1-1)[(\Gs_1/\Gs_2+1)V+x]
$$
at the boundary of the inclusion. Since the only function which has zero real part at the boundary of a
closed curve, and which is analytic in the interior, is an imaginary constant, we deduce that $(\Gs_1/\Gs_2+1)V+x$ is constant around the boundary of
the inclusion and hence constant inside, {\it i.e.}, in the inclusion $-\Grad V= \begin{bmatrix} k \\ 0 \end{bmatrix}$
with  $k=1/(\Gs_1/\Gs_2+1)$.

The harmonic conjugate to $V$ inside the inclusion is then $-ky$ which can
be identified with $(\Gs_2/\Gs_1)W$ (to within an additive constant). Then from the first condition in \eq{AE12}
it follows that (to within an additive constant) $V'$ takes the value $-ky$ around the
boundary of the inclusion and hence in its interior
too,  {\it i.e.}, in the inclusion $-\Grad V'= \begin{bmatrix} 0 \\ k \end{bmatrix}$. Hence the uniform field
attainability condition is met, and the upper bound is attained.

We summarize our findings as a theorem.
\begin{theorem}
Provided the body $\GO$ consists of inclusions of phase 1 completely surrounded by phase 2 then the upper bound
is attained if and only if there exist potentials $V$ and $V'$, which are linear combinations of the
potentials $V_1$ and $V_2$, such that $V-iV'+x$ is an analytic function of $z=x+iy$ in phase 2.
\end{theorem}

Similarly we have the following theorem for the lower bound.
\begin{theorem}
Provided the body $\GO$ consists of inclusions of phase 2 completely surrounded by phase 1 then the lower bound
is attained if and only if there exist potentials $V$ and $V'$, which are linear combinations of the
potentials $V_1$ and $V_2$, such that $V-iV'+x$ is an analytic function of $z=x+iy$ in phase 1.
\end{theorem}

\section{Asymptotic bounds for small volume fraction}

Suppose that the phase 1 occupies a region $\omega \subset \Om$ satisfying
 \beq
 \mbox{dist} (\omega, \p\Om) \ge c
 \eeq{AB1}
for some $c>0$. The purpose of this section is to compare the bounds \eq{LB39} and \eq{UB10} with the bounds obtained in \cite{CV03} when the volume $|\omega|$ of $\omega$ tends to $0$.

Let $q$ be a function on $L^2(\p\Om)$ satisfying $\int_{\p\Om} q=0$. Let $V$ be the solution to
 \beq
 \left\{
 \begin{array}{l}
 \Grad \cdot \Gs \Grad V=0  \quad  \mbox{in } \Om, \\
 \nm
 \dis \Gs \frac{\p V}{\p \Bn} = q \quad \mbox{on } \p \Om, \quad (\int_{\p\Om} V =0),
 \end{array}
 \right.
 \eeq{AB2}
and let $U$ be the solution to \eq{AB2} with $\Gs$ replaced with $\Gs_2$. It is proved in \cite{CV022} that given a sequence $\omega_n$ satisfying \eq{AB1} and such that $|\omega_n| \to 0$
there is a subsequence still denoted $\omega_n$, a probability measure $d\mu$ supported in the set $\{ x ~|~ \mbox{dist} (x, \p\Om) \ge c \}$, and a (pointwise) polarization tensor field $M(\Bx)$ such that if $V_n$ is the solution to \eq{AB2} when $\omega=\omega_n$, then
 \beq
 V_n (\Bx) - U(\Bx) = -|\omega_n| \int_{\Om} \Grad U(\Bz) \cdot M(\Bz) \Grad_z N(\Bx, \Bz) d\mu(\Bz) + o(|\omega_n|), \quad \Bx \in \p\Om,
 \eeq{AB3}
where $N(\Bx,\Bz)$ is the Neumann function for $\Om$, {\it i.e.}, $U$ is given by
 \beq
 U(\Bz)= \int_{\p\Om} N(\Bx, \Bz) q(\Bx) ds(\Bx).
 \eeq{AB4}
Note that we have absorbed a factor of $\Gs_1-\Gs_2$ into the definition of $M$ given by Capdeboscq and Vogelius to be consistent with the conventional definition of polarization tensors.

Let $V'_n$ be the solution to
 \beq
 \left\{
 \begin{array}{l}
 \Grad \cdot \Gs \Grad V=0  \quad  \mbox{in } \Om, \\
 V = V^0 \quad \mbox{on } \p \Om
 \end{array}
 \right.
 \eeq{AB5}
with $\omega=\omega_n$ and $U'$ be the solution to \eq{AB5} with $\Gs$ replaced with $\Gs_2$. Then we have
 \beq
 \Gs_2 \frac{\p V'_n}{\p \Bn} (\Bx) - \Gs_2 \frac{\p U'}{\p \Bn} (\Bx) = |\omega_n| \int_{\Om} \Grad U'(\Bz) \cdot M(\Bz) \Grad_z \frac{\p }{\p \Bn_\Bx} G(\Bx, \Bz) d \mu(\Bz) + o(|\omega_n|), \quad \Bx \in \p\Om,
 \eeq{AB6}
where $G(\Bx, \Bz)$ is the Green function for $\Om$, {\it i.e.}, $U'$ is given by
 \beq
 U'(\Bz)= \int_{\p\Om} \frac{\p}{\p \Bn_\Bx} G(\Bx, \Bz) V^0(\Bx) ds(\Bx).
 \eeq{AB7}

To see \eq{AB6} let us define the Neumann-to-Dirichlet (NtD) map $\GL_\Gs$ by
 \beq
 \GL_\Gs [q]:= V|_{\p\Om}
 \eeq{AB8}
where $V$ the solution to \eq{AB2}. Let $\GL_{\Gs_2}$ be the NtD map when $\Gs$ is replaced with $\Gs_2$. Observe that
because of \eq{AB4}, we have
 \begin{align}
 \int_{\Om} \Grad U(\Bz) \cdot M(\Bz) \Grad_z N(\Bx, \Bz) d \mu(\Bz) &= \int_{\p\Om} \left[ \int_{\Om} \Grad_z N(\By, \Bz)\cdot M(\Bz) \Grad_z N(\Bx, \Bz) d \mu(\Bz) \right] q(\By) ds(\By) \nonumber \\
 & := K[q](\Bx). \label{AB9}
 \end{align}
So \eq{AB3} can be rewritten as
 \beq
 \GL_\Gs[q]= \GL_{\Gs_2}[q] - |\Go_n| K[q] + o(|\Go_n|).
 \eeq{AB10}
Then the Dirichlet-to-Neumann map $\GL_\Gs^{-1}$ is given by
 \beq
 \GL_\Gs^{-1} = (I - |\Go_n| \GL_{\Gs_2}^{-1} K)^{-1} \GL_{\Gs_2}^{-1} + o(|\Go_n|) = \GL_{\Gs_2}^{-1} + |\Go_n| \GL_{\Gs_2}^{-1} K \GL_{\Gs_2}^{-1} + o(|\Go_n|).
 \eeq{AB11}
Observe that
 \beq
 \GL_{\Gs_2}^{-1} [N(\cdot, \Bz)](\Bx) = \frac{\p}{\p \Bn_\Bx} G(\Bx, \Bz), \quad \Bx \in \Om, \ \ \Bz \in \Om.
 \eeq{AB12}
In fact,
 \beq
 \int_{\p\Om} \GL_{\Gs_2}^{-1} [N(\cdot, \Bz)](\Bx) V^0(\Bx) =  \int_{\p\Om} N(\Bx, \Bz) \Gs_2 \frac{\p U'}{\p\Bn} (\Bx) = U'(\Bz) \quad \mbox{for all } \Bz \in \Om,
 \eeq{AB13}
and hence \eq{AB12} follows. We now obtain \eq{AB6} from \eq{AB11}.

Let $U_1(\Bx)= -\Gs_2^{-1} x$ and $U_2(\Bx)= -\Gs_2^{-1} y$, and let $V_j$ be the solution to \eq{AB2} with $q=q_j$ for $j=1,2$, where $q_j=\Gs_2 \frac{\p U_j}{\p \Bn}=-n_j$. Then, we have
 \beq
[\BGs_N^{-1}]_{ij}= a_{ij}= \frac{1}{|\Om|} \int_{\p\Om} V_i q_j = \frac{1}{|\Om|} \int_{\p\Om} (V_i - U_i) q_j + \Gs_2^{-1} \delta_{ij}
 \eeq{AB14}
where $\delta_{ij}$ is Kronecker's delta. Since
 \beq
 \int_{\p\Om} N(\Bx, \Bz) q_j(\Bx) ds(\Bx) = U_j(\Bz),
 \eeq{AB15}
we have from \eq{AB3} that
 \beq
 \frac{1}{|\Om|} \int_{\p\Om} (V_i - U_i) q_j = - |\Go_n| \frac{1}{\Gs_2^2 |\Om|} \int_{\Om} M_{ij} (\Bx) d\mu(x) + o(|\Go_n|) .
 \eeq{AB16}
Thus we have
 \beq
 \BGs_N^{-1}= - f_1 \Gs_2^{-2} M + \Gs_2^{-1} I + o(f_1)
 \eeq{AB17}
where
 \beq
 M:= \int_{\Om} M_{ij} (\Bx) d\mu(\Bx).
 \eeq{AB18}
We then have
 \begin{align}
 \frac{\Tr\BGs_N^{-1} - 2 \Gs_1^{-1}}{\det\BGs_N^{-1} - \Gs_1^{-2}}
 &= \frac{-\frac{f_1}{\Gs_2^2} \Tr M + \frac{2}{\Gs_2} - \frac{2}{\Gs_1} + o(f_1)}{-\frac{f_1}{\Gs_2^3} \Tr M + \frac{1}{\Gs_2^2} - \frac{1}{\Gs_1^2} + o(f_1)} \nonumber \\
 &= \frac{ \frac{2(\Gs_1-\Gs_2)}{\Gs_1 \Gs_2} \left[ 1 -\frac{f_1 \Gs_1}{2\Gs_2(\Gs_1 - \Gs_2)} \Tr M \right] + o(f_1)}{ \frac{(\Gs_1^2 -\Gs_2^2)}{\Gs_1^2 \Gs_2^2} \left[ 1 -\frac{f_1 \Gs_1^2}{\Gs_2(\Gs_1^2 - \Gs_2^2)} \Tr M \right] + o(f_1)} \nonumber \\
 &= \frac{2 \Gs_1 \Gs_2}{\Gs_1 + \Gs_2} \left[ 1 + \frac{f_1 \Gs_1}{2\Gs_2(\Gs_1 + \Gs_2)} \Tr M \right] + o(f_1), \label{AB19}
 \end{align}
and hence
 \beq
 \frac{(\Gs_1 + \Gs_2)}{\Gs_1 (\Gs_1 - \Gs_2)} \left[ \frac{\Tr \BGs_N^{-1}  - 2/\Gs_1}{\det \BGs_N^{-1} - 1/\Gs_1^2} -
 \frac{2 \Gs_1 \Gs_2}{\Gs_1 + \Gs_2} \right] = \frac{f_1 \Gs_1}{(\Gs_1^2 - \Gs_2^2)} \Tr M + o(f_1).
 \eeq{AB20}
The lower bound \eq{HS6} now reads
 \beq
 \frac{f_1 \Gs_1}{(\Gs_1^2 - \Gs_2^2)} \Tr M \le f_1,
 \eeq{AB21}
or equivalently
 \beq
 \frac{\Gs_1}{(\Gs_1^2 - \Gs_2^2)} \Tr (I-\Gs_2\BGs_N^{-1}) \le f_1
 \eeq{AB22}
up to $o(f_1)$ terms by \eq{AB17}.

We now consider the upper bound. Let $U_1(\Bx)=-x$ and $U_2(\Bx)=-y$, and let $V_i$ be the solution to \eq{AB5} with $V^0=U_i$ on $\p\Om$ for $i=1,2$. Then, defining $q_j=\Gs_2 \frac{\p V_j}{\p \Bn}$ on $\p\GO$,
we have
 \beq
[\BGs_D]_{ij}= a_{ij}= \frac{1}{|\Om|} \int_{\p\Om} V_i^0 q_j = \frac{1}{|\Om|} \int_{\p\Om} V_i^0 (q_j - \Gs_2\frac{\p U_j}{\p \Bn}) + \Gs_2 \delta_{ij}
 \eeq{AB23}
One can use \eq{AB6} and the fact that
 \beq
 \int_{\p\Om} \frac{\p}{\p \Bn_{\Bx}} G(\Bx, \Bz) V_i^0(\Bx) d\Bx= U_i(\Bz)
 \eeq{AB24}
to derive that
 \beq
 \BGs_D= f_1 M + \Gs_2 I + o(f_1)=\BGs_N + o(f_1).
 \eeq{AB25}
Thus we obtain
 \beq
 \frac{1}{\Gs_2} - \frac{\Tr\BGs_D - 2 \Gs_2}{\det\BGs_D - \Gs_2^2} = \frac{f_1}{\Gs_2^2} \frac{\det M}{\Tr M} + o(f_1).
 \eeq{AB26}
Since $\det M = (\Tr M^{-1})^{-1} \Tr M$, \eq{HS10} reads
 \beq
 f_1 \le \frac{f_1 (\Gs_1+\Gs_2)}{\Gs_2(\Gs_1-\Gs_2)} (\Tr M^{-1})^{-1} ,
 \eeq{AB27}
or equivalently
 \beq
 f_1 \le \frac{(\Gs_1+\Gs_2)}{(\Gs_1-\Gs_2)} (\Tr (-I+\Gs_2^{-1}\BGs_D)^{-1})^{-1}
 \eeq{AB28}
up to $o(f_1)$ terms.

By \eq{AB17} and \eq{AB25}, we have
 \beq
 -I+ \Gs_2^{-1} \BGs_D = I - \Gs_2\BGs_N^{-1}
 \eeq{AB29}
modulo $o(f_1)$. Hence by putting \eq{AB22} and \eq{AB28} together, we have
 \beq
 \frac{\Gs_1 \Gs_2}{(\Gs_1^2 - \Gs_2^2)} \Tr (I-\Gs_2\BGs_N^{-1}) \le f_1 \le \frac{(\Gs_1+\Gs_2)}{(\Gs_1-\Gs_2)} (\Tr (I-\Gs_2\BGs_N^{-1})^{-1})^{-1}
 \eeq{AB30}
modulo $o(f_1)$, where $\BGs_N^{-1}$ is determined from the boundary measurements with special Neumann conditions, via \eq{AB14}.
We emphasize that these asymptotic bounds for small volume fraction were found in \cite{CV022, CV03}. From \eq{AB29} we also have the bounds
\beq \frac{\Gs_1 \Gs_2}{(\Gs_1^2 - \Gs_2^2)} \Tr (\Gs_2^{-1}\BGs_D-I) \le f_1 \le \frac{(\Gs_1+\Gs_2)}{(\Gs_1-\Gs_2)} (\Gs_2^{-1}\BGs_D-I)^{-1})^{-1}
\eeq{AB35}
modulo $o(f_1)$, where $\BGs_D$ is obtained from the boundary measurements with special Dirichlet conditions.

It is interesting to observe that the translation bounds also yield the Lipton bounds for the polarization tensor: We obtain from \eq{AB21} and \eq{AB27} that
\beq
\Tr M \le \frac{(\Gs_1^2 - \Gs_2^2)}{\Gs_1}  \quad\mbox{and}\quad \Tr (M^{-1}) \le
\frac{(\Gs_1+\Gs_2)}{\Gs_2(\Gs_1-\Gs_2)} .
\eeq{AB36}
We refer to \cite{book, book2, milton} for properties of polarization tensors. If the phase 1 is an inclusion (or a cluster of inclusions) of the form
\beq
D= \ep B + \Bz
\eeq{AB37}
where $\ep$ is a small parameter representing the diameter of $D$, $B$ is a reference domain containing $0$, and
$\Bz$ indicates the location of $D$ inside $\GO$, then $M_{ij}=|B|^{-1} M(B)$ (a constant matrix)
and $d\mu = \lim_{n \to \infty} |\omega_n|^{-1} \chi(\omega_n) d\Bx=\Gd(\Bx-\Bz)d\Bx$. Here $M(B)$ is the polarization tensor associated with $B$. Therefore we have
$M=|B|^{-1} M(B)$, and hence
\beq
\Tr(M(B))\leq |B| \frac{(\Gs_1^2 - \Gs_2^2)}{\Gs_1},
\eeq{AB38}
and the lower bound is given by
\beq
\Tr (M(B)^{-1})\leq\frac{\Gs_1+\Gs_2}{\Gs_2(\Gs_1-\Gs_2) |B|}.
\eeq{AB39}
The bounds in \eq{AB38} and \eq{AB39} were obtained by Lipton \cite{Lipton93} and later by Capdeboscq-Vogelius \cite{CV022, CV03} in a more general setting. They can also easily be derived
from the bounds of Lurie and Cherkaev \cite{lucherk82} and Tartar and Murat \cite{mutar85,tar85} using the observation made by Milton \cite{milt81} that the low volume fraction limit
of bounds on effective tensors of periodic arrays of well-separated inclusions yield bounds on polarization tensors.
We also mention that if the lower bound in \eq{AB39} is attained for $B$ and $B$ is simply connected, then $B$ is an ellipse.
This was known as the P\'olya-Szeg\"o conjecture and resolved by Kang-Milton \cite{KM06, KM07} (see also a review paper \cite{Kang09}).

\section{Numerical results}

\subsection{Forward solutions}
\begin{figure}[htb]
\begin{center}
\epsfig{figure=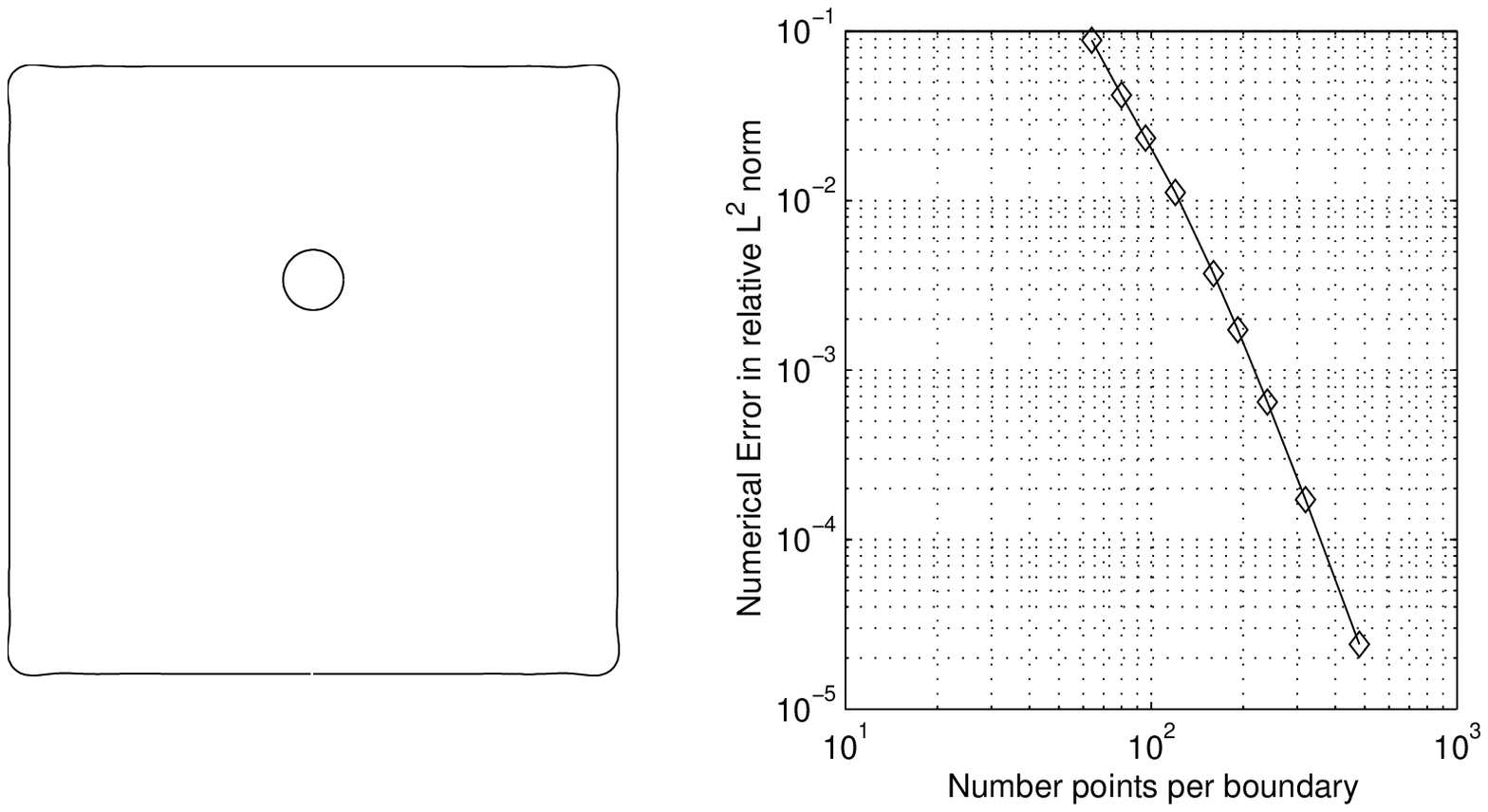,width=10cm}
\end{center}
\caption{Convergence error of the forward solver with 64-480
discretization points. The solid line represents the
convergence error of $V$ on $\p\Om$ for the Neumann problem.}\label{conv.1}
\end{figure}

\begin{figure}[htb]
\begin{center}
\epsfig{figure=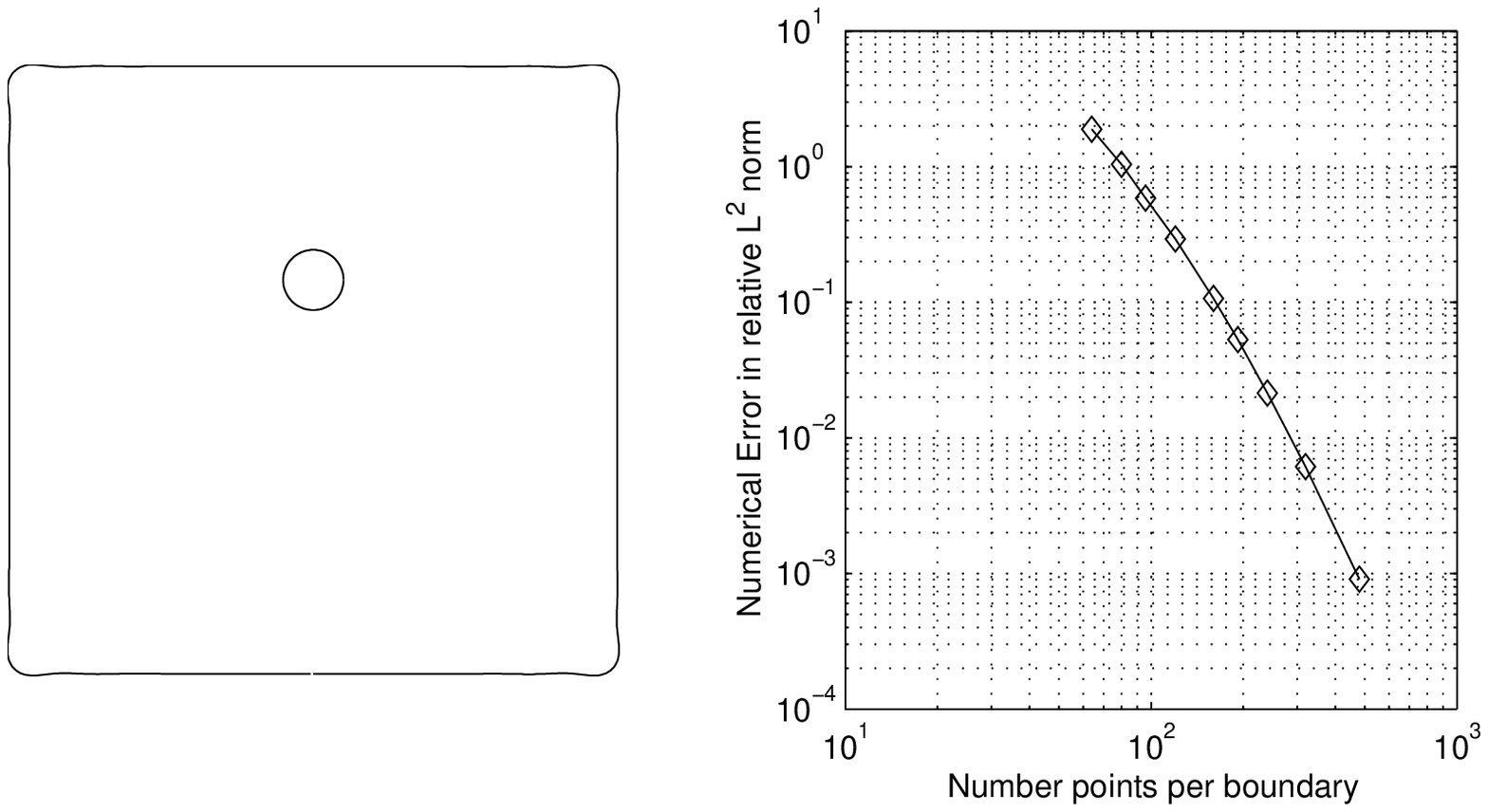,width=10cm}
\end{center}
\caption{Convergence error of the forward solver with 64-480
discretization points. The solid line represents the
convergence error of $\frac{\p V}{\p\Bn}$ for the Dirichlet
problem.}\label{conv.2}
\end{figure}

We implement an integral equation solver in FORTRAN in order to
generate forward solutions of the Neumann and Dirichlet problems of the equation $\Grad \cdot \Gs \Grad V=0$ in $\Om$ when $D$ is an inclusion and $\Gs = \Gs_1 \chi(D) + \Gs_2 \chi(\Om\setminus D)$. We set $\Gs_2=1$ throughout this section. We
compute the forward solutions $V$ with
$N=64,80,96,120,160,192,240,320$ and $480$ equi-spaced points on $\p D$ and $N$
points on $\p\Omega$. And then they are computed with the solutions on the finer discretization of $N=960$. Figure \ref{conv.1} shows the convergence of a forward solver for the Neumann problem as a function of discretization points, $N$, and Figure \ref{conv.2} for the Dirichlet problem, with $\Gs_1=10$

\subsection{Numerical Experiments}

We perform numerical simulations to judge the performance of the bounds when relevant parameters are varying. Parameters under consideration are the conductivity contrast $\Gs_1/\Gs_2$, the volume fraction $f_1$, and the distance between the inclusion and $\p\Om$. We also investigate the role of boundary data in deriving bounds.

We use boundary data of special forms; $q_1=-\begin{bmatrix} 1 \\ 0 \end{bmatrix}\cdot\Bn$
and $q_2=-\begin{bmatrix} 0 \\ 1 \end{bmatrix}\cdot\Bn$ as Neumann data for the lower bound, and
$V_1=-\begin{bmatrix}
  1 \\
  0
\end{bmatrix}\cdot \Bx$ and $V_2=-\begin{bmatrix}
  0 \\
 1
\end{bmatrix}\cdot \Bx$ as Dirichlet data for the upper bound, in all examples except examples \ref{7.4} and \ref{7.5}. Thus except in these examples, the bounds correspond
to those derived by Milton \cite{milt11}.

Let
\begin{equation}
L(\Gs_1):= \frac{(\Gs_1 + \Gs_2)}{\Gs_1 (\Gs_1 - \Gs_2)} \left[ \frac{\Tr A - 2b/\Gs_1}{\det A - b^2/\Gs_1^2} -
 \frac{2 \Gs_1 \Gs_2}{\Gs_1 + \Gs_2} \right],
\end{equation}
denote the lower bound on $f_1$
and let
\begin{equation}
U(\sigma_1) := \frac{\Gs_2 (\Gs_1+\Gs_2)}{(\Gs_1-\Gs_2)} \left[ \frac{1}{\Gs_2} - \frac{\Tr A - 2b' \Gs_2}{\det A - b'^2 \Gs_2^2} \right]
\end{equation}
denote the upper bound on $f_1$.

\begin{figure}[htbp]
\begin{center}
\epsfig{figure=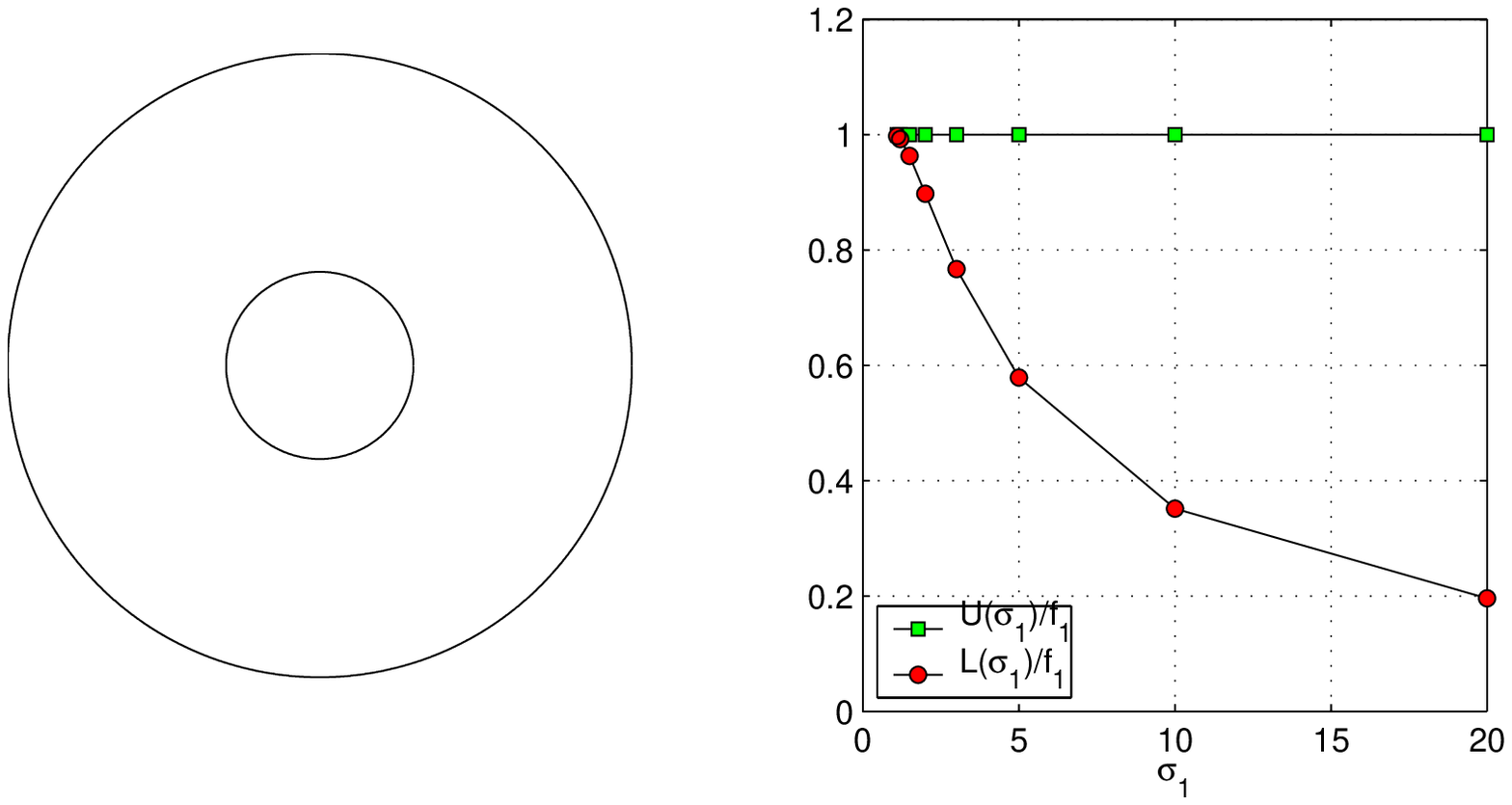,width=8cm}\\
\epsfig{figure=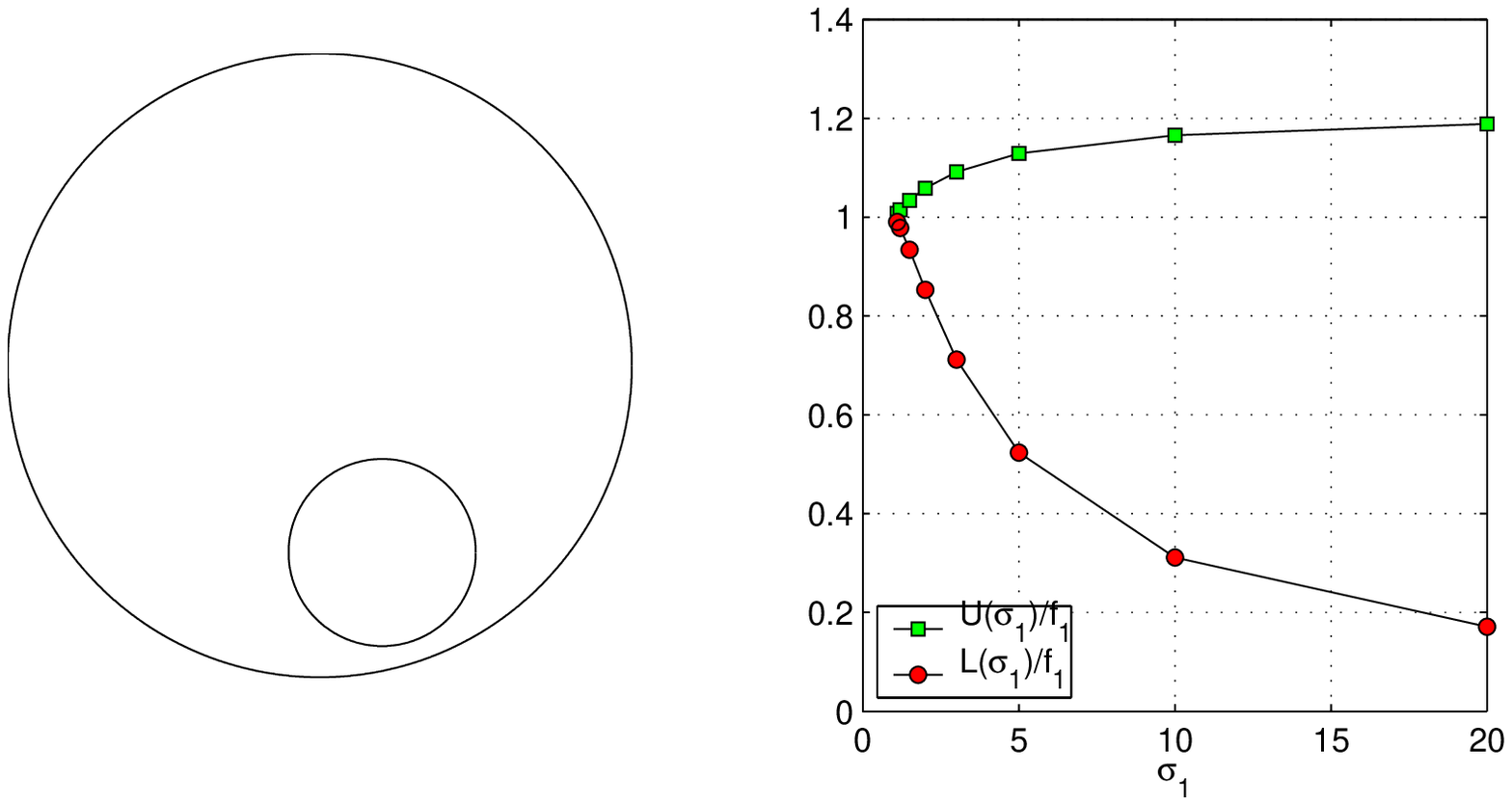,width=8cm}\vskip 0.5cm
\begin{tiny}
\title{first diagram\ \ \ \ \ \ \ \ \ \ \ \ \ \ \ \ \ \ \ \ \ \ \ \ \ \ \ \ second diagram}\\
\begin{tabular}{|c ||c |c|}\hline
  $\sigma_1$ & $L(\sigma_1)/f_1$
 & $U(\sigma_1)/f_1$ \\\hline
1.1&0.9979&1.0000\\
1.2&0.9925&1.0000\\
1.5&0.9635&1.0000\\
2  &0.8979&1.0000\\
3  &0.7673&1.0000\\
5  &0.5787&1.0000\\
10 &0.3518&1.0000\\
20 &0.1958&1.0000\\\hline
\end{tabular}\hskip 0.5cm
\begin{tabular}{|c ||c |c|}\hline
  $\sigma_1$ & $L(\sigma_1)/f_1$
 & $U(\sigma_1)/f_1$ \\\hline
1.1&0.9904&1.0077\\
1.2&0.9783&1.0149\\
1.5&0.9340&1.0337\\
2  &0.8532&1.0583\\
3  &0.7115&1.0917\\
5  &0.5237&1.1287\\
10 &0.3113&1.1659\\
20 &0.1710&1.1889\\\hline
\end{tabular}
\end{tiny}
\end{center}
 \caption{The bounds with increasing $\sigma_1$ when the inclusion
is a disk and $\GO$ is a circle, and $f_1=0.09$. We take Neumann and Dirichlet data of the special
forms. The second and third columns are graphs of the same data; the third column is with a log-scale for the $\sigma_1$-axis. The values for the bounds are given in the table.}\label{LU_DYDY1}
\end{figure}

\begin{figure}[htbp]
\begin{center}
\epsfig{figure=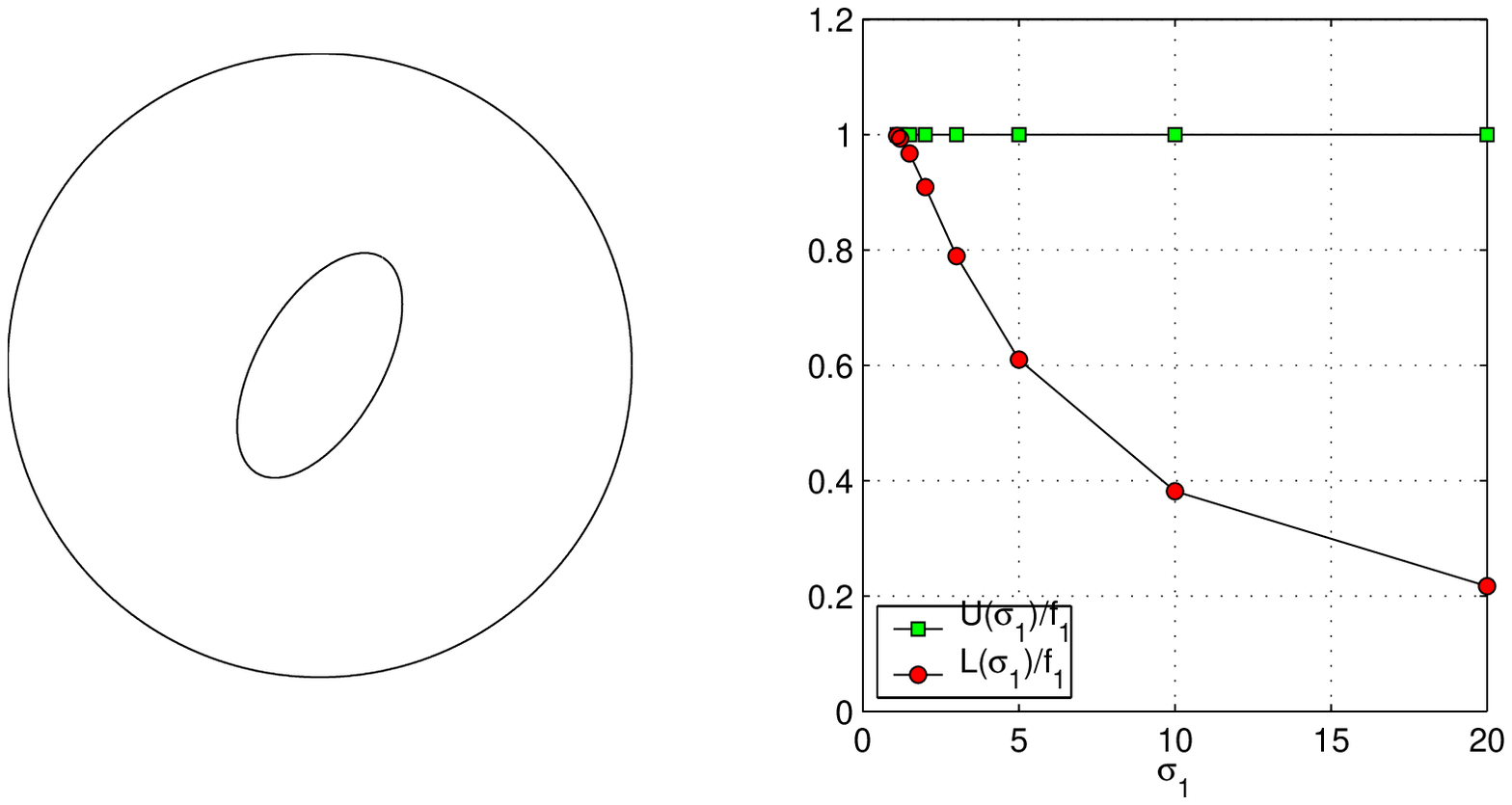,width=8cm}\\
\epsfig{figure=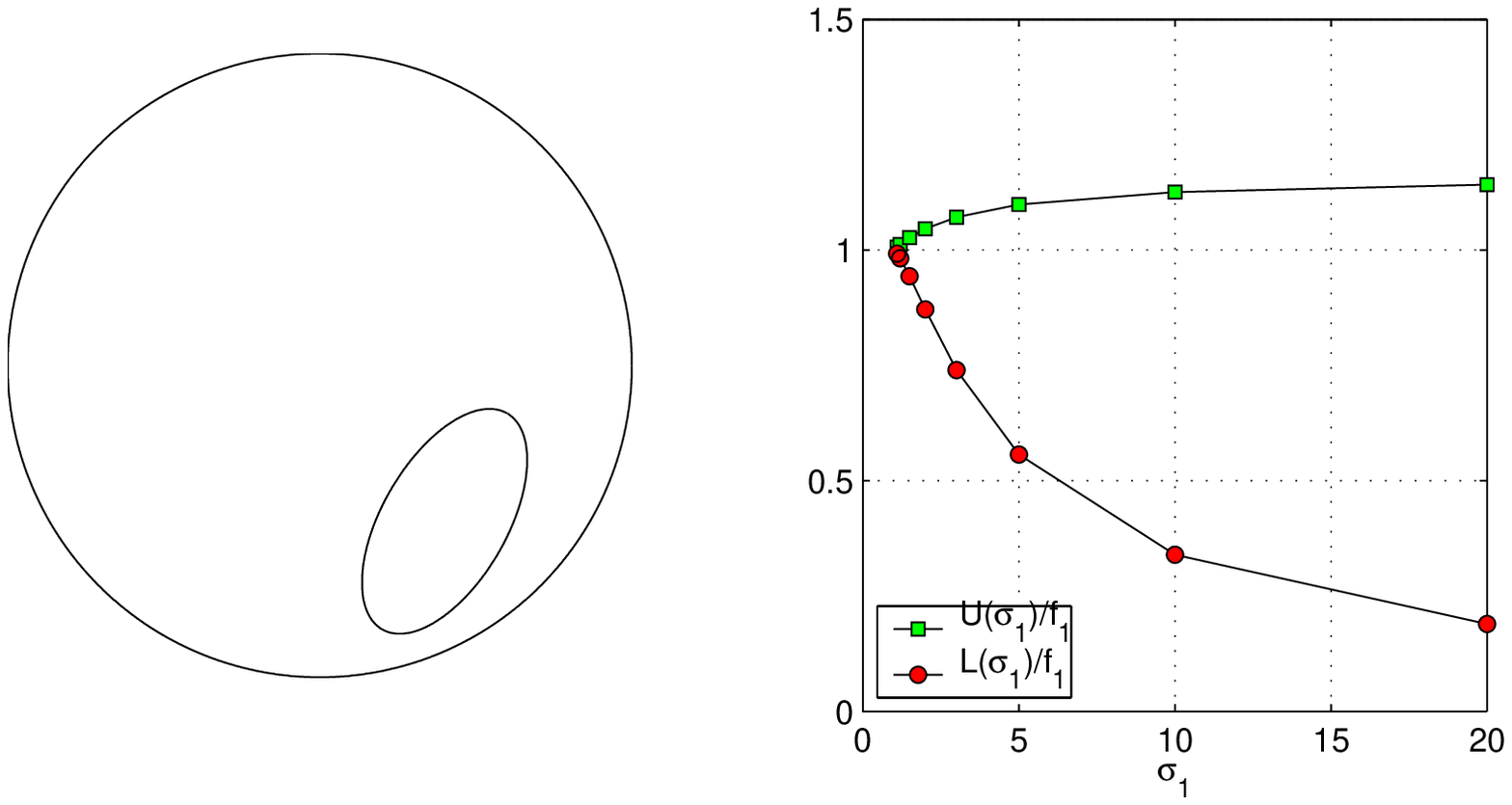,width=8cm}\vskip 0.5cm
\begin{tiny}
\title{first diagram\ \ \ \ \ \ \ \ \ \ \ \ \ \ \ \ \ \ \ \ \ \ \ \ \ \ \ \ second diagram}\\
\begin{tabular}{|c ||c |c|}\hline
  $\sigma_1$ & $L(\sigma_1)/f_1$
 & $U(\sigma_1)/f_1$ \\\hline
1.1&0.9982&1.0000\\
1.2&0.9934&1.0000\\
1.5&0.9677&1.0000\\
2  &0.9091&1.0001\\
3  &0.7895&1.0001\\
5  &0.6099&1.0001\\
10 &0.3818&1.0002\\
20 &0.2170&1.0002\\\hline
\end{tabular}\hskip 0.5cm
\begin{tabular}{|c ||c |c|}\hline
  $\sigma_1$ & $L(\sigma_1)/f_1$
 & $U(\sigma_1)/f_1$ \\\hline
1.1&0.9921 &1.0062\\
1.2&0.9819&1.0119\\
1.5&0.9435&1.0268\\
2  &0.8712&1.0459\\
3  &0.7396&1.0714\\
5  &0.5569&1.0988\\
10 &0.3395&1.1257\\
20 &0.1896&1.1420\\\hline
\end{tabular}
\end{tiny}
\end{center}
\caption{The bounds with increasing $\sigma_1$ when the inclusion is an ellipse and $\Omega$ is a circle and $f_1=0.08$. We take Neumann and Dirichlet data of the special
forms. The second and third columns are graphs of the same data; the third column is with a log-scale for the $\sigma_1$-axis. The values for the bounds are given in the table.}\label{LU_DYDY3}
\end{figure}

\begin{figure}[htbp]
\begin{center}
\epsfig{figure=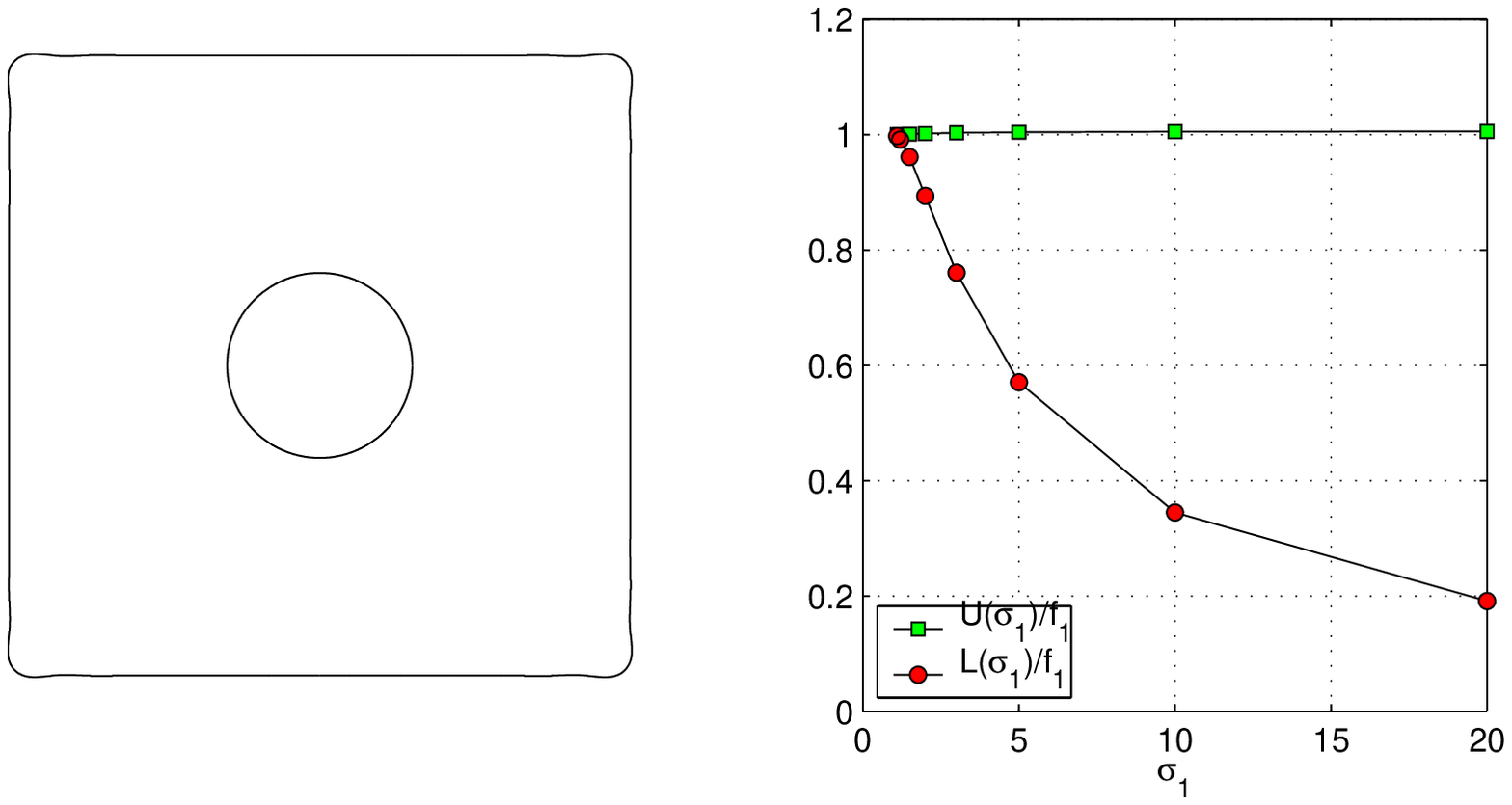,width=8cm}\\
\epsfig{figure=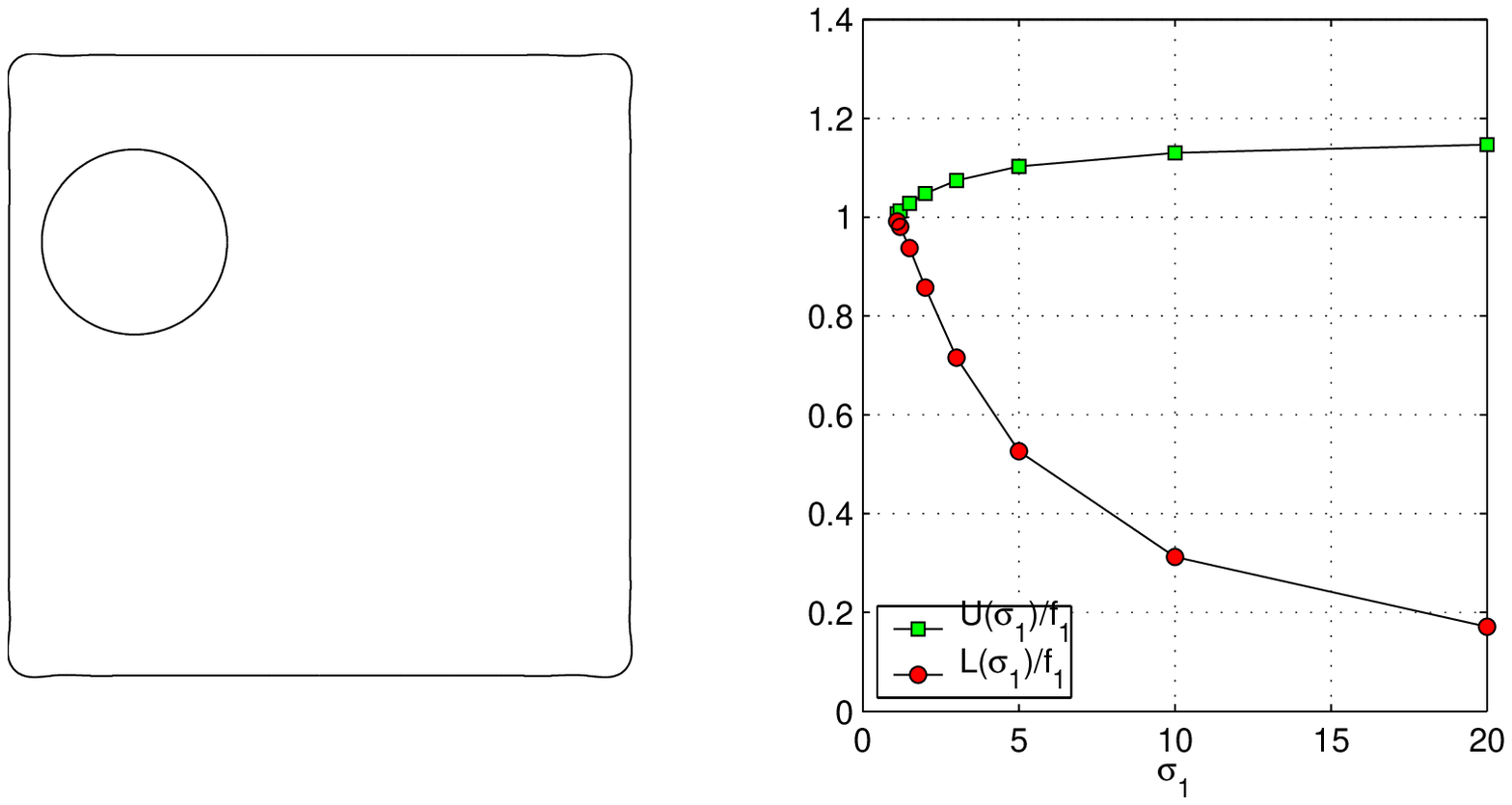,width=8cm}\vskip 0.5cm
\begin{tiny}
\title{first diagram\ \ \ \ \ \ \ \ \ \ \ \ \ \ \ \ \ \ \ \ \ \ \ \ \ \ \ \ second diagram}\\
\begin{tabular}{|c ||c |c|}\hline
  $\sigma_1$ & $L(\sigma_1)/f_1$
 & $U(\sigma_1)/f_1$ \\\hline
1.1&0.9976&1.0003\\
1.2&0.9917&1.0006\\
1.5&0.9614&1.0013\\
2  &0.8939&1.0022\\
3  &0.7608&1.0033\\
5  &0.5708&1.0044\\
10 &0.3449&1.0054\\
20 &0.1912&1.0060\\\hline
\end{tabular}\hskip 0.5cm
\begin{tabular}{|c ||c |c|}\hline
  $\sigma_1$ & $L(\sigma_1)/f_1$
 & $U(\sigma_1)/f_1$ \\\hline
1.1&0.9915&1.0065\\
1.2&0.9803&1.0125\\
1.5&0.9376&1.0281\\
2  &0.8576&1.0480\\
3  &0.7155&1.0744\\
5  &0.5262&1.1027\\
10 &0.3122&1.1302\\
20 &0.1713&1.1468\\\hline
\end{tabular}
\end{tiny}
\end{center}
\caption{The bounds with increasing $\sigma_1$ when the inclusion is a
disk and $\GO$ is a square, and $f_1=0.0699$.  We take Neumann and Dirichlet data of the special
forms. The second and third columns are graphs of the same data; the third column is with a log-scale for the $\sigma_1$-axis. The values for the bounds are given in the table.}\label{LU_DYDY2}
\end{figure}

\begin{figure}[htbp]
\begin{center}
\epsfig{figure=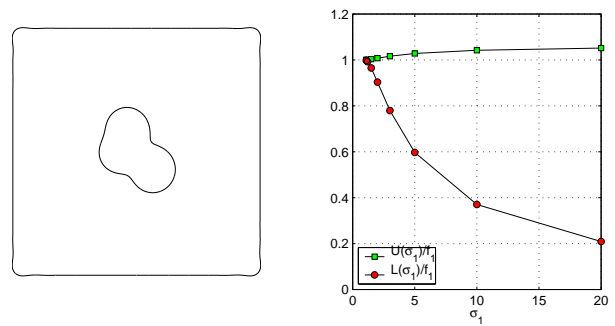,width=8cm}
\end{center}
\caption{The bounds with increasing $\sigma_1$ when the inclusion is
not a disk or an ellipse and $\GO$ is a square, and $f_1=0.0673$. We take Neumann and Dirichlet data of the special
forms. The second and third columns are graphs of the same data; the third column is with a log-scale for the $\sigma_1$-axis.}\label{LU_DYDN}
\end{figure}

\begin{figure}[htbp]
\begin{center}
\epsfig{figure=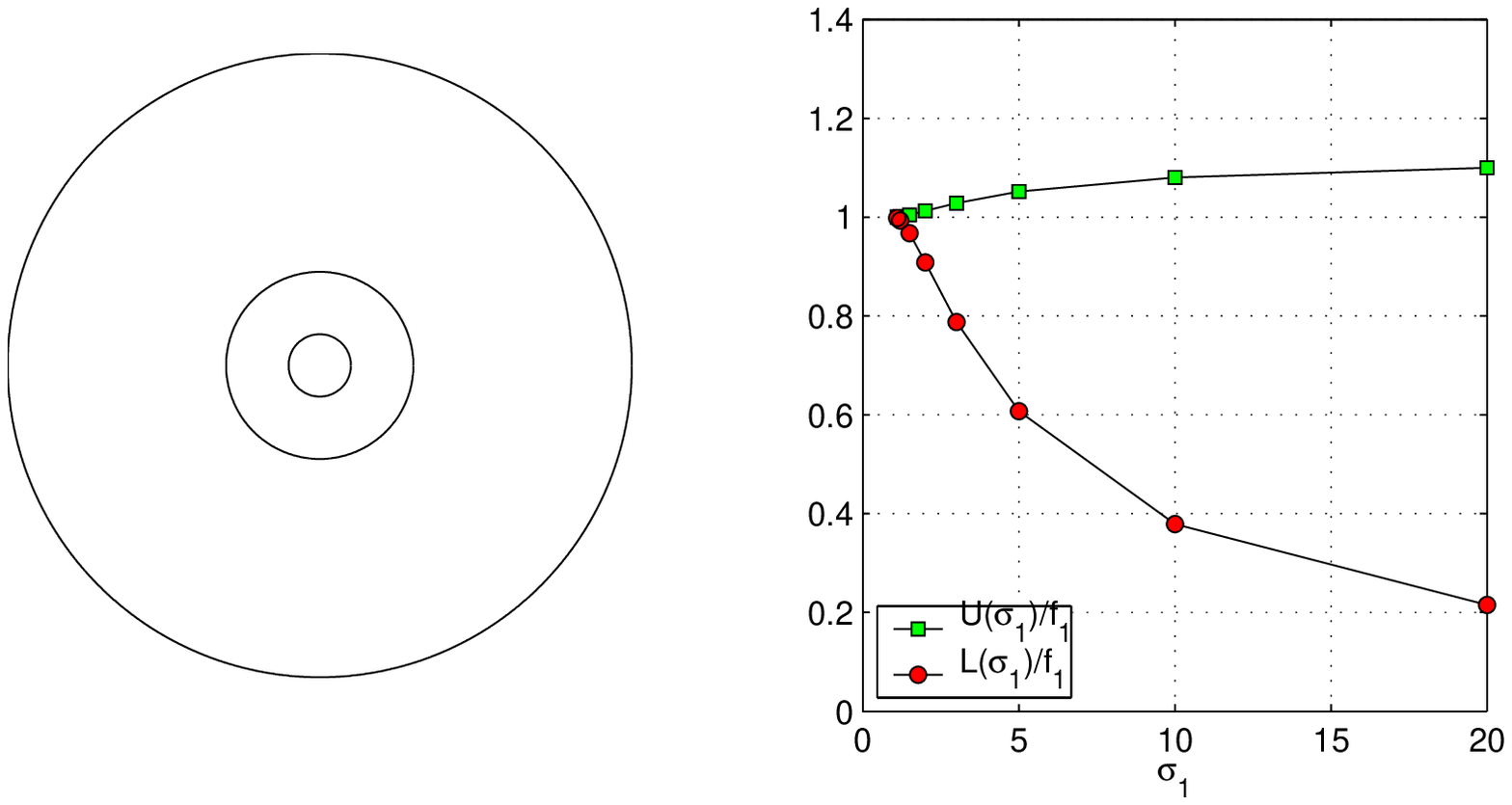,width=8cm}\\
\epsfig{figure=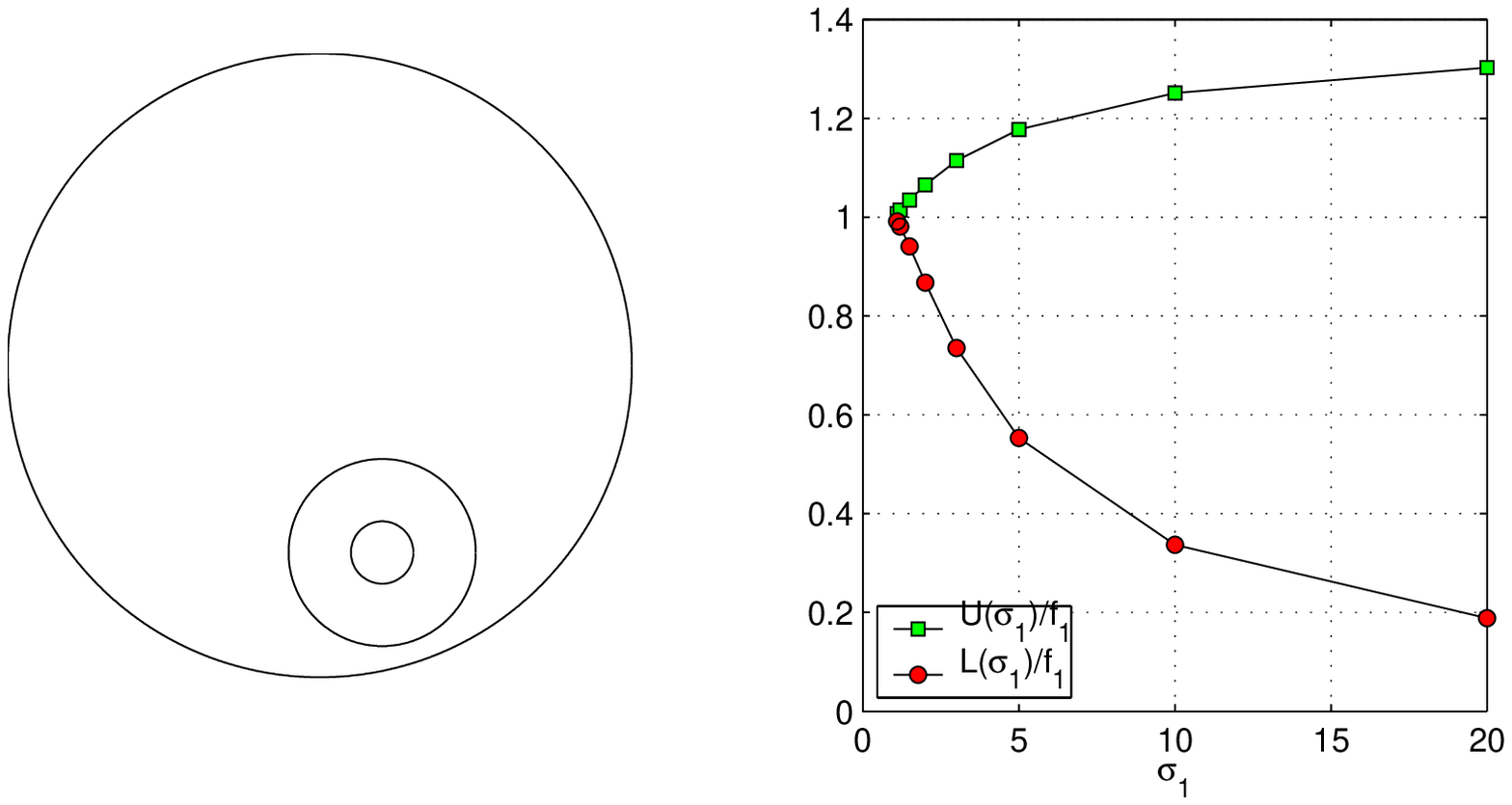,width=8cm}\\
\epsfig{figure=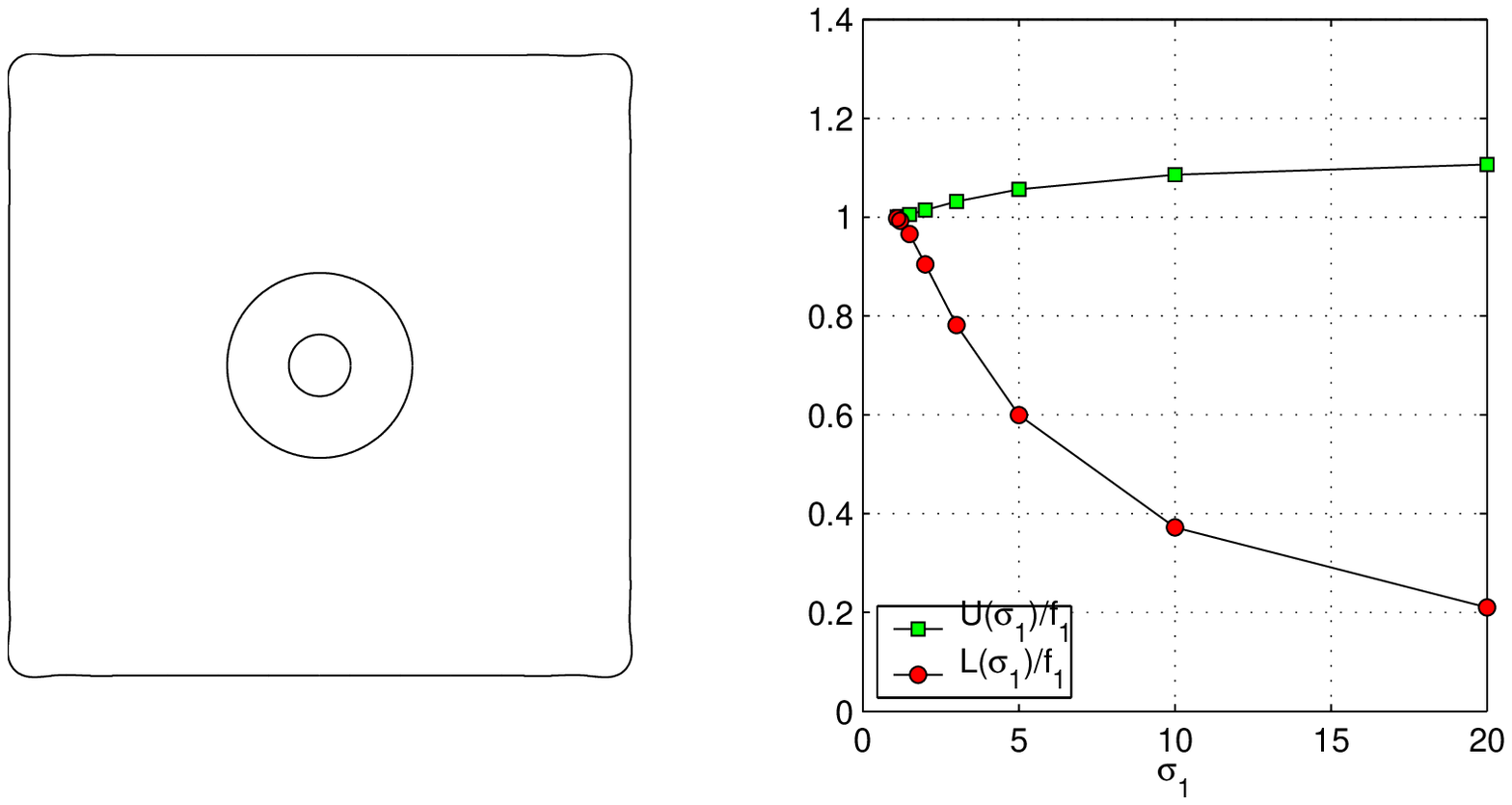,width=8cm}\\
\epsfig{figure=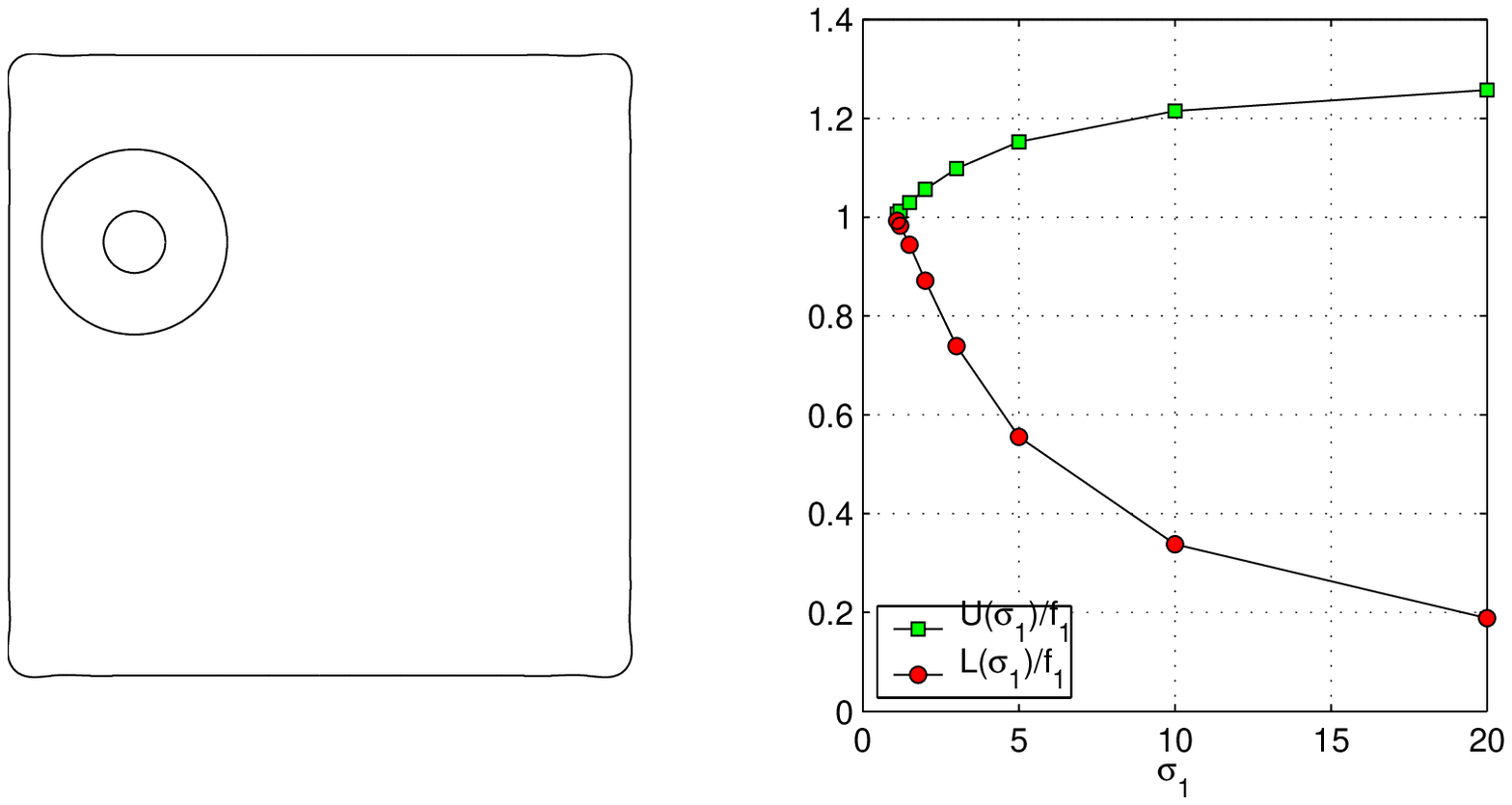,width=8cm}
\end{center}
\caption{The bounds with increasing $\sigma_1$ when the inclusion is an
annulus. We take Neumann and Dirichlet data of the special
forms. The second and third columns are graphs of the same data; the third column is with a log-scale for the $\sigma_1$-axis.}\label{ALU}
\end{figure}

\begin{Exa} {\bf (variation of $\Gs_1$)}.
We compute the bounds changing $\sigma_1$, keeping $\Gs_2=1$, when
the inclusion is a disk or an ellipse inside a disk or a rectangle (with corners rounded).
Figures \ref{LU_DYDY1}, \ref{LU_DYDY3}, and \ref{LU_DYDY2} show the numerical results.
Figure \ref{LU_DYDN} is when the inclusion is simply connected and of general shape. Figure \ref{ALU}
is when the inclusion is not simply connected.
The results show that the lower bound deteriorates seriously as the conductivity ratio $\Gs_1$ increases while the upper bounds are relatively good even with large $\Gs_1$.
\end{Exa}

\begin{figure}[htbp]
\begin{center}
\epsfig{figure=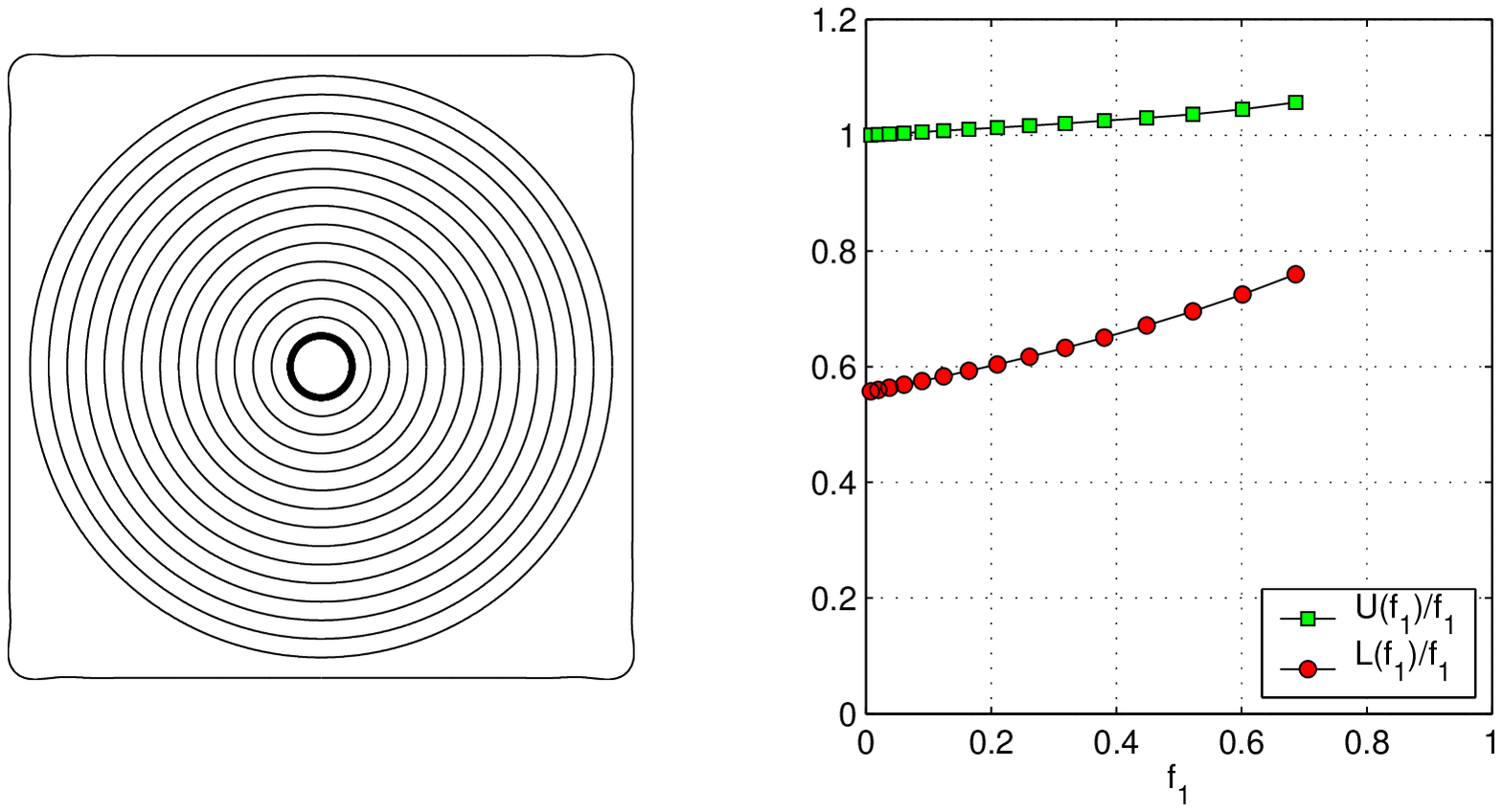,width=7cm}\\
\epsfig{figure=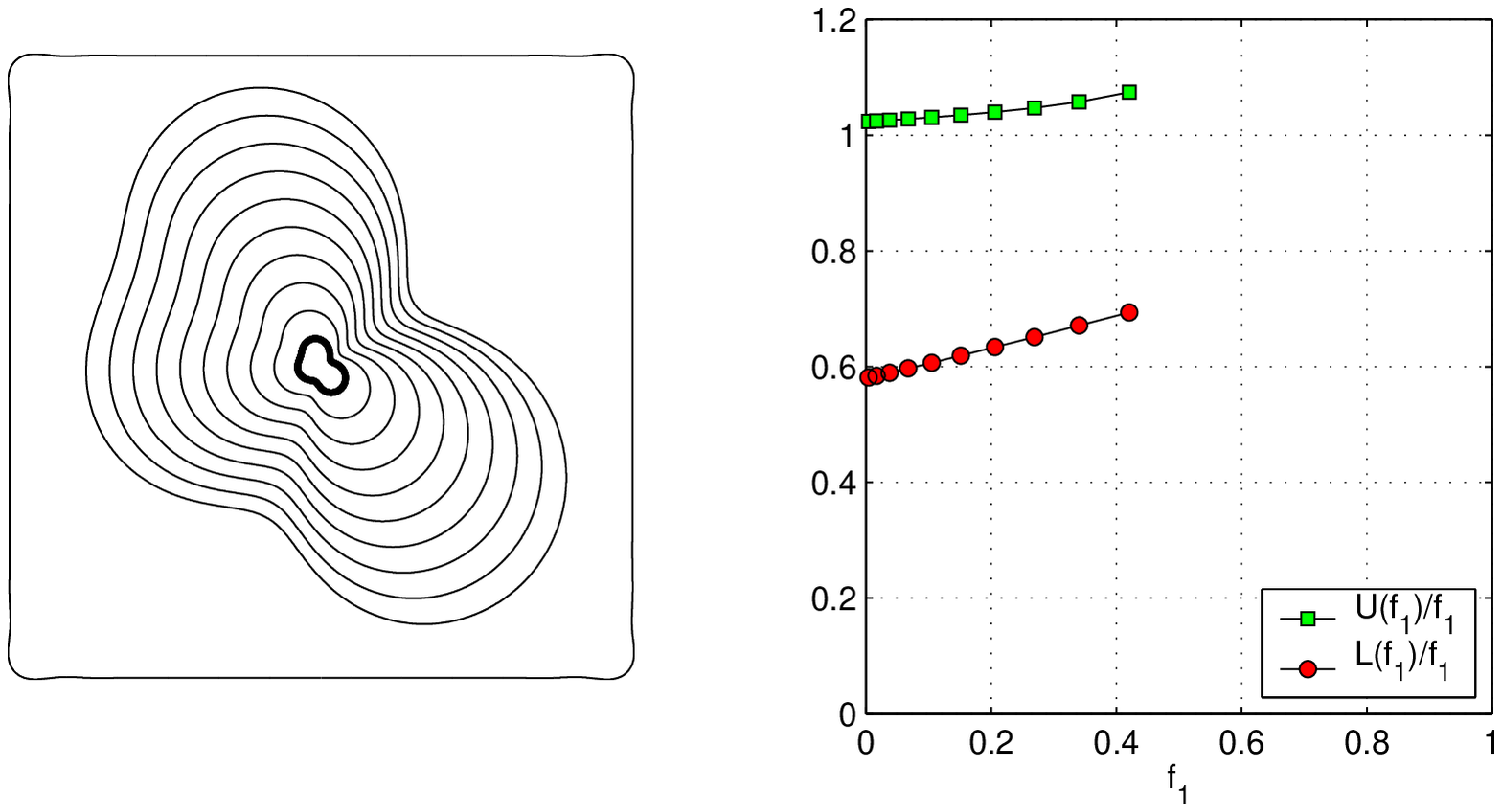,width=7cm}
\end{center}
\caption{$\sigma_1=5$. The bounds changing the volume fraction. We
take Neumann and Dirichlet data   with the special forms.
}\label{LU_area}
\end{figure}

\begin{Exa}{\bf (variation of $f_1$)}.
We compute the bounds for various volume fractions.
Figure \ref{LU_area} shows the numerical results. It clearly shows that the lower bound works better for higher volume fractions.
\end{Exa}

\begin{figure}[htbp]
\begin{center}
\epsfig{figure=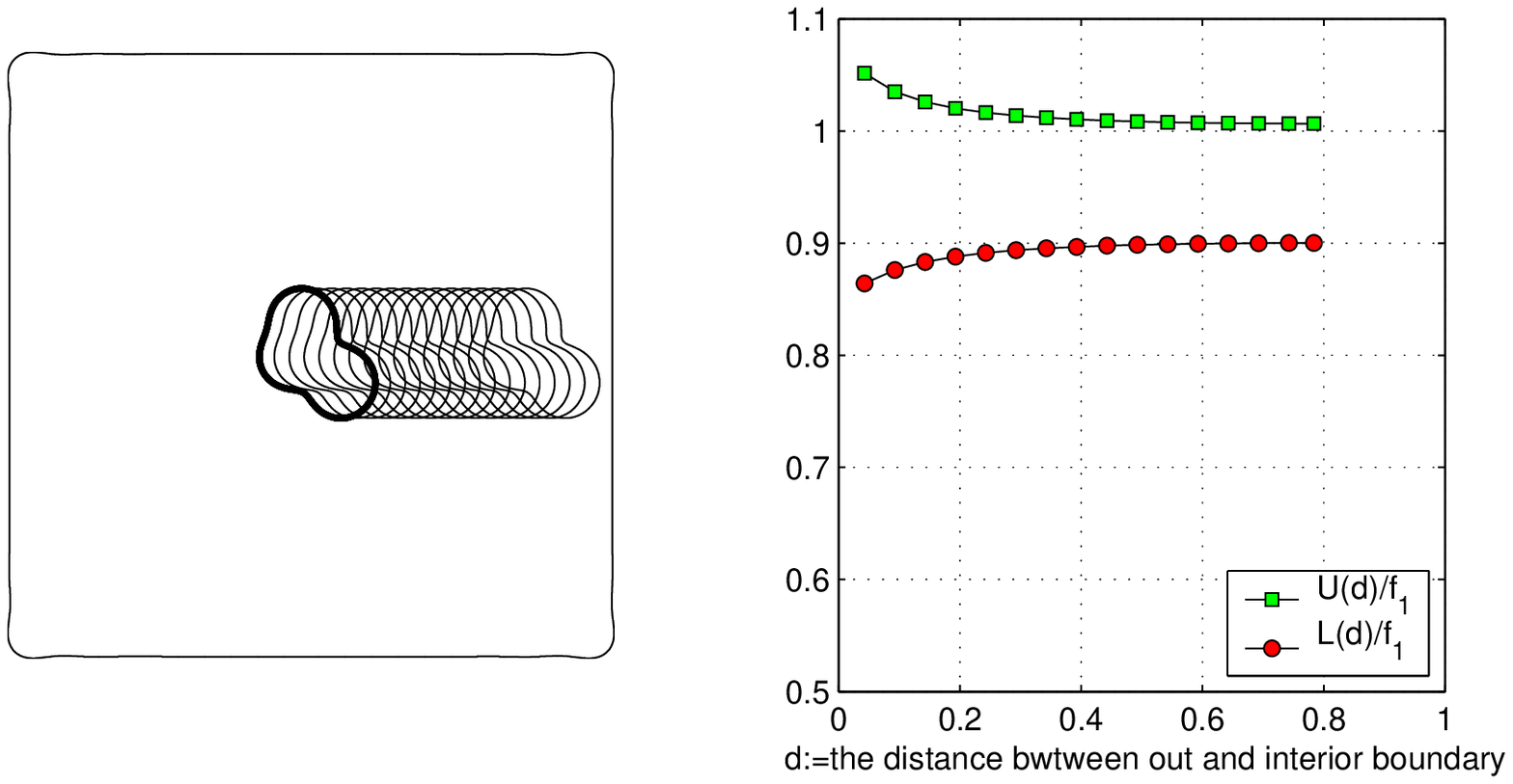,width=8cm}
\end{center}
\caption{$\sigma_1=2$ and $f_1=0.0262$. The bounds changing the distance between
the inclusion and $\p\Om$. We take Neumann and Dirichlet
data of the special forms. }\label{LU_dist1}
\end{figure}

\begin{Exa}{\bf (variation of distance from $\p\Om$)}.
We compute the lower and upper bounds changing
the distance between the inclusion and $\p\Om$. Figure \ref{LU_dist1}
shows the numerical results when $\sigma_1=2$. It shows that the further the inclusion is from $\p\Om$, the better bounds are.
\end{Exa}

\begin{figure}[htbp]
\begin{center}
\epsfig{figure=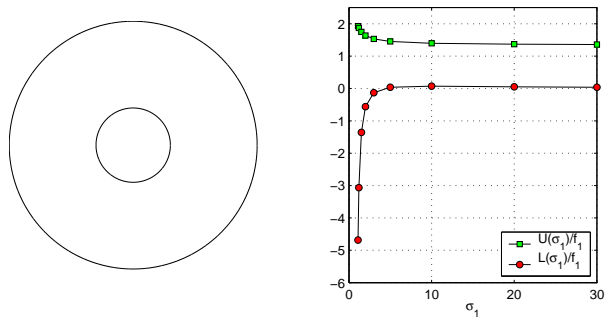,width=8cm}
\end{center}
\caption{The bounds changing $\sigma_1$ in case that we take
Neumann data $g_1=-\nu_1-\nu_1\nu_2$ and $g_2=-\nu_2-\nu_1\nu_2$.
Also we take Dirichlet data $V_1=-x-xy$ and $V_2=-y-xy$.
}\label{LU_gDYDY}
\end{figure}

\begin{Exa}\label{7.4} {\bf (boundary data)}.
In the example we compute the bounds using other boundary data. We use as Neumann data for the lower bound
$q_1=-n_1 - n_1 n_2$ and $q_2=-n_2 - n_1 n_2$, and as Dirichlet data for the upper bound
$V_1=-x-xy$ and $V_2=-y-xy$. Figure \ref{LU_gDYDY} shows that the special boundary data work much better.
\end{Exa}

\begin{figure}[htbp]
\begin{center}
\epsfig{figure=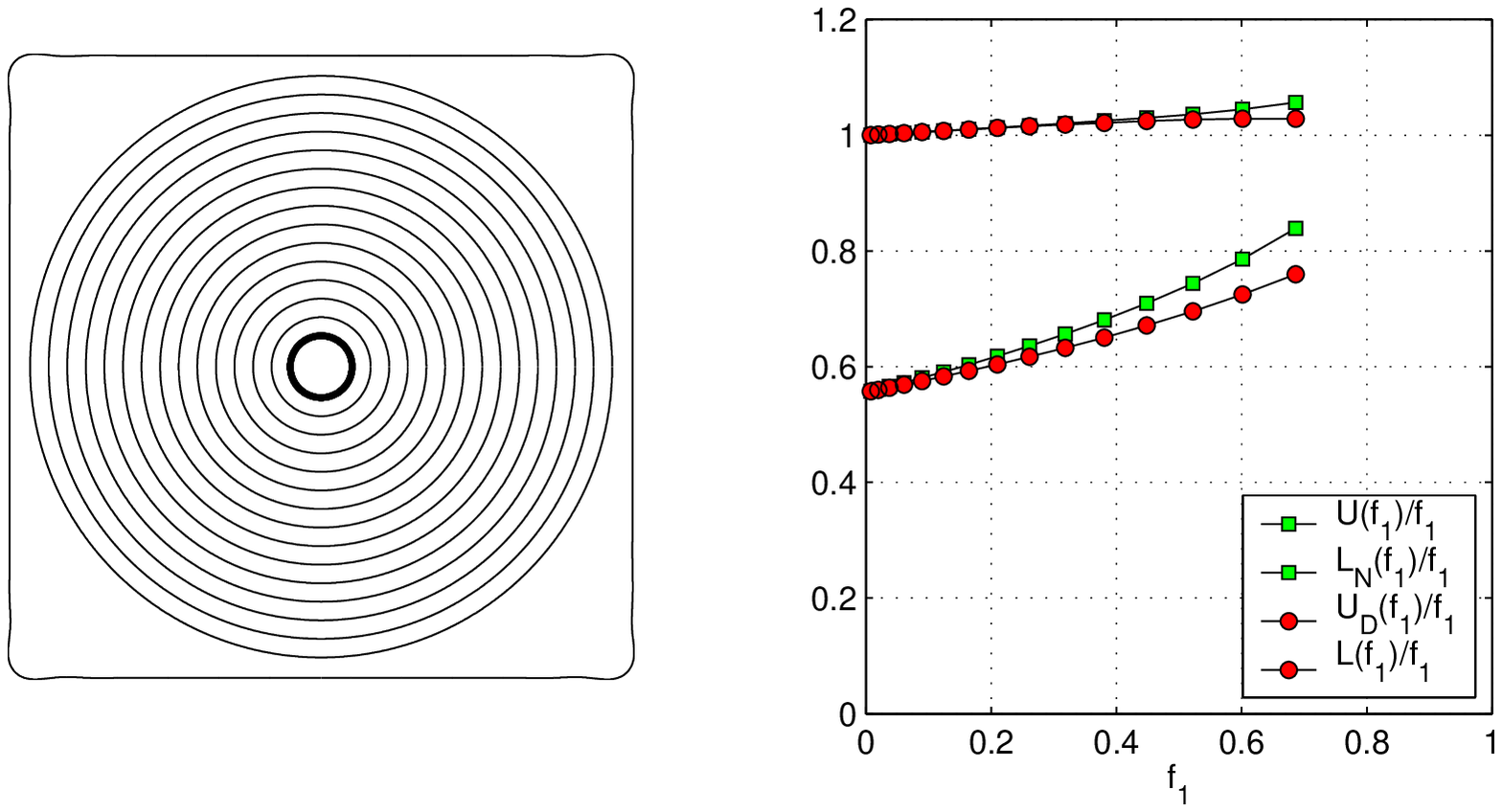,width=8cm}
\epsfig{figure=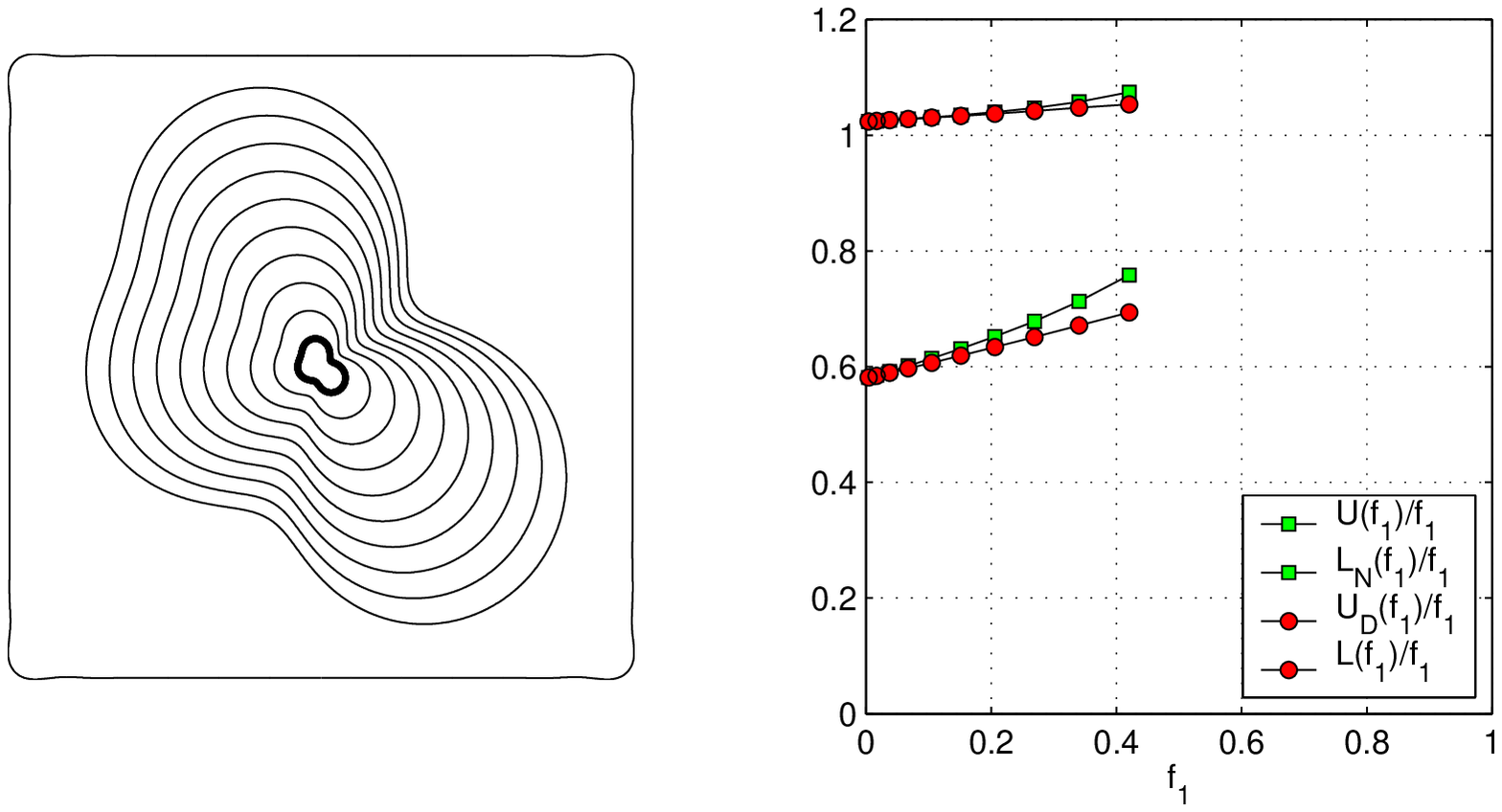,width=8cm}
\end{center}
\caption{$\sigma_1=5$. $L_N(f_1)$ is the lower bound using the Neumann data corresponding to the special Dirichlet data and $U_D(f_1)$ is the upper bound using the Dirichlet data corresponding to the special Neumann data.
}\label{LU_same}
\end{figure}

\begin{Exa}\label{7.5} When we use special Neumann data $q_1=-n_1$ and $q_2=-n_2$, then a pair of Dirichlet data are measured on $\p\Om$. We may use this data to compute the upper bound using the formula \eq{UB19}. Likewise, we may use the measured Neumann data corresponding to the Dirichlet data $V_1=-x$ and $V_2=-y$ to compute the lower bound using the formula \eq{LB45}. Figure \ref{LU_same} shows numerical results when the volume fraction varies. In this example it clearly shows that bounds using the measured data are better than those using the given data.
\end{Exa}

\section{Construction of E$_\Om$-inclusions}

Following the method outlined in section 23.9 of \cite{milton} we look for a simply connected inclusion inside which the field is uniform for some boundary condition assigned on the outer boundary. More precisely, we look for an inclusion $E$ contained in a domain $\Om$ (bounded or unbounded) such that $-\nabla V$ is uniform inside $E$, where $V$ is the solution to
 \beq
 \left\{
 \begin{array}{l}
 \Grad \cdot \Gs \Grad V=0  \quad  \mbox{in } \Om, \\
 V=V^0 \quad \mbox{on } \p\Om
 \end{array}
 \right.
 \eeq{CE1}
for some boundary data $V^0$ with $\Gs=\Gs_1 \chi(E) + \Gs_2 \chi(\Om \setminus E)$ ($\Gs_1 \neq \Gs_2$). We may suppose, without loss of generality, that $\Be=-\Grad V = (-1, 0)^T$. We also suppose that the coordinates have been positioned and scaled so that $y_{\mbox{max}}=1$ and $y_{\mbox{min}}=-1$, where $y_{\mbox{max}}=\max \{ y ~ | ~(x,y) \in E \mbox{ for some } x \}$ and $y_{\mbox{min}}=\min \{ y ~| ~(x,y) \in E \mbox{ for some } x \}$. Let $W$ be a harmonic conjugate of $V$ in $\Om \setminus \overline{E}$ so that $V+iW$ is an analytic function of $z=x+iy$ in $\Om \setminus \overline{E}$. Then we have
 \beq
 V=x, \quad W= \frac{\Gs_1}{\Gs_2} y, \quad \mbox{on } \p E.
 \eeq{CE2}
Define new potentials $u$ and $v$ by
 \beq
 u+ i v = \frac{i(V+iW-z)}{(1-\Gs_1/\Gs_2)}.
 \eeq{CE3}
Then, $u+ i v$ is still an analytic function of $z=x+iy$ in $\Om \setminus \overline{E}$, and on $\p E$
 \beq
 u = \frac{-W+y}{1-\Gs_1/\Gs_2} = y, \quad v = \frac{V-x}{1-\Gs_1/\Gs_2} = 0.
 \eeq{CE4}

Now assume $u+ i v$ is a univalent function of $z=x+iy$ inside $\Om \setminus \overline{E}$, and consider $x+iy$ as an analytic function of $u+ i v$ (hodograph transformation). Because of \eq{CE4}, the image of $\p E$ by $u+ i v$ is the slit $S= [y_{\mbox{min}}, y_{\mbox{max}}] = [-1,1]$ on the $u$-axis, and $y= u$ on $S$.

The problem is now to construct a function $z=x+iy=f(u+iv)$ such that
\begin{itemize}
\item[(i)] $f$ is analytic and univalent in $U \setminus S$ for some neighborhood $U$ of $S$,
\item[(ii)] $\Imag f=u$ on $S$,
\item[(iii)] $\Real f|_{+} - \Real f|_{-} >0$ on $S$ except at $\pm 1$ where it is $0$.
\end{itemize}
Here $|_{+}$ and $|_{-}$ indicate the limit from above and below $S$, respectively.

One can see that the conditions (i), (ii), and (iii) guarantee that $f$ maps $U \setminus S$ onto $\Om \setminus \overline{E}$ for a simply connected domain $E$ and $\Om$ a domain containing $\overline{E}$. In fact, (ii) and (iii) imply that $f$ maps $S$ onto $\p E$ and the orientation is preserved. Since $f$ is conformal, it maps $U \setminus S$ to outside $\overline{E}$.

We have the following lemma for univalence.

\begin{lemma} \label{lemma21}
Let $\gamma$ be a simple closed curve which consists of two curves $\gamma^+$ and $\gamma^-$. Let $U$ be an open neighborhood of $S$ and let $B_1(\Gd)$ and $B_{-1}(\Gd)$ be open balls of radius $\Gd$ centered at $w=1$ and $w=-1$
respectively. Let $f$ be an analytic function in $U\setminus S$ which maps $U\setminus S$ outside $\gamma$ of the form
 \beq
 f(w)=iw + g(w)
 \eeq{CE5}
where $\Imag g=0$ on $S$. Suppose that the mapping $u \mapsto \lim_{v \to 0^+} f(u+iv)$ is one-to-one from $S$ onto $\gamma^+$, and $u \mapsto \lim_{v \to 0^+} f(u-iv)$ is one-to-one from $S$ onto $\gamma^-$. If there is $\delta>0$ such that $f$ is univalent in $B_1(\Gd)\setminus S$ and in $B_{-1}(\Gd)\setminus S$, then there is an open neighborhood $U_0$ of $S$ such that $f$ is univalent in $U_0 \setminus S$.
\end{lemma}

\proof Let $\varphi(z)=(z+\frac{1}{z})/2$ for $|z| \ge 1$. $\varphi$ maps $|z|>1$ onto $\mathbb{C} \setminus S$ and $|z|=1$ onto $S$. Let $G(z)= g(\varphi(z))$. Since $\Imag G(z)=0$ on $|z|=1$, $G$ can be extended so that it is analytic in $1-\ep <|z| < 1+ \ep$ for some $\ep>0$. Let $F(z)= f(\varphi(z))$. Then $F$ is analytic in $1-\ep <|z| < 1+ \ep$ and univalent in the neighborhoods of $z=1$ and $z=-1$.
Moreover, $F$ is one-to-one from $|z|=1$ onto $\gamma$. We claim that $F$ is univalent in $1-\ep_0 <|z| < 1+ \ep_0$ for some $\ep_0>0$. In fact, if not, then for each $n$ there are $z_{1,n}$ and $z_{2,n}$ such that $1- \frac{1}{n} < |z_{j,n}| < 1+ \frac{1}{n}$, $z_{1,n} \neq z_{2,n}$, and $F(z_{1,n})=F(z_{2,n})$. For $j=1,2$, the sequence $z_{j,n}$ has a subsequence which converges to a point on $|z|=1$, say $z_j$. Since $F$ is one-to-one on $|z|=1$, $z_1=z_2$.
But this implies that $F'(z_1)=0$, where
\beq F'(z)=f'(\varphi(z))\varphi'(z)=[i+g'(\varphi(z))](1-z^{-2})/2,
\eeq{CM0}
and since $g'(\varphi(z_1))$ is real we conclude that $z_1=1$ or $z_1=-1$ which is contradiction since $F$ is univalent in the neighborhoods of these points.
Thus $F$ is univalent in $1-\ep_0 <|z| < 1+ \ep_0$ for some $\ep_0>0$. This completes the proof. \qed

We now construct $f$ satisfying (i), (ii), and (iii) using conformal mappings.
Let $w=u+iv$ and define
 \beq
 g(w)=f(w)-iw
 \eeq{CM1}
so that $\Imag g=0$ on $S$. Let
 \beq
 \xi = \frac{1-w}{1+w},
 \eeq{CM2}
which maps $S$ onto the positive real axis. Let $\zeta=\sqrt{\xi}$ with the branch cut along the positive real axis and define
 \beq
 F(\zeta) = g \left( \frac{1-\zeta^2}{1+\zeta^2} \right).
 \eeq{CM3}
Then $\Imag F=0$ on the whole real axis. Thus, by defining $F(\zeta^*)=F(\zeta)^*$, where $*$ denotes the complex conjugate, $F$ can be extended as an analytic function in a tubular neighborhood of the real axis. Moreover, since $g$ is analytic in a neighborhood of $-1$ except the part of the slit and the bilinear transform $\zeta$ maps a neighborhood of $-1$ onto outside a compact set, $F$ must be analytic in $\mathbb{C} \setminus (K \cup K^*)$ where $K$ is a compact set in the upper half plane and $K^*$ is its symmetric part with respect to the real axis, {\it i.e.}, $K^*=\{z^* ~|~ z \in K\}$. $F$ satisfies
\begin{itemize}
\item[(i)$^\prime$] $F$ is analytic in $\mathbb{C} \setminus (K \cup K^*)$ for a compact set $K$ in the upper half plane.
 \item[(ii)$^\prime$] $\Imag F=0$ on the real axis,
\item[(iii)$^\prime$] $F(\zeta) - F(-\zeta) >0$  for real positive $\zeta$.
\end{itemize}

The function $f$ is now given by
\beq
 f(w)=iw + F\left( \sqrt{\frac{1-w}{1+w}}. \right)
\eeq{CM4}
Note that $y=u$ on the slit and hence $\p E$ is given by
 \beq
 x = F\left( \pm \sqrt{\frac{1-y}{1+y}} \right).
 \eeq{CM5}

In addition to (i)$^\prime$, (ii)$^\prime$, and (iii)$^\prime$, $F$ needs to be univalent inside a suffiently small ball around the origin, and outside a sufficiently large ball. The first condition is satisfied if
$F'(0)\ne 0$. Since $F$ maps $\infty$ to a point in $\mathbb{C}$, $F$ being analytic and univalent outside a sufficiently large ball has the series expansion
 \beq
 F(\zeta)= \sum_{j=0}^\infty \frac{\beta_j}{\zeta^j}
 \eeq{CM6}
as $\zeta \to \infty$, where $\beta_1 \ne 0$ (and $\beta_1$ is real and positive from conditions (ii)$^\prime$, and (iii)$^\prime$).

We make a record of these conditions:
\begin{itemize}
\item[(iv)$^\prime$]
The derivative $F'(0)$ is non-zero, and $F(\zeta)$ has the asymptotic expansion
 \beq
 F(\zeta)= \beta_0 + \frac{\beta_1}{\zeta} + O(|\zeta|^{-2}) \quad\mbox{as } |\zeta| \to \infty,
 \eeq{CM7}
where $\Gb_1$ is real and positive.
\end{itemize}

Good candidates for functions satisfying (i)$^\prime$, (ii)$^\prime$, and (iv)$^\prime$ are rational functions of the form
 \beq
 F(\zeta)= \sum_{\Ga=1}^n \left[ \frac{b_\Ga}{\zeta-a_\Ga} + \frac{b_\Ga^*}{\zeta-a_\Ga^*} \right] + c
 \eeq{CM8}
where the $a_\Ga$'s are complex numbers with positive imaginary parts, the $b_\Ga$'s are complex numbers, $c$ is a real number,
and
\beq  \sum_{\Ga=1}^n \Real (b_\Ga)>0,\quad \sum_{\Ga=1}^n \Real (b_\Ga/a_\Ga^2)\ne 0.
\eeq{CM8a}
To ensure that (iii)$^\prime$ is satisfied we require that the function
\beq F(\zeta) - F(-\zeta)=2\zeta\sum_{\Ga=1}^n \left[ \frac{b_\Ga}{\zeta^2-a_\Ga^2} + \frac{b_\Ga^*}{\zeta^2-(a_\Ga^*)^2} \right]
\eeq{CM8b}
has no real roots aside from $\zeta=0$. (The sign of the inequality in (iii)$^\prime$ is guaranteed by the positivity of $\beta_1$.)

\begin{figure}[htbp]
\begin{center}
\epsfig{figure=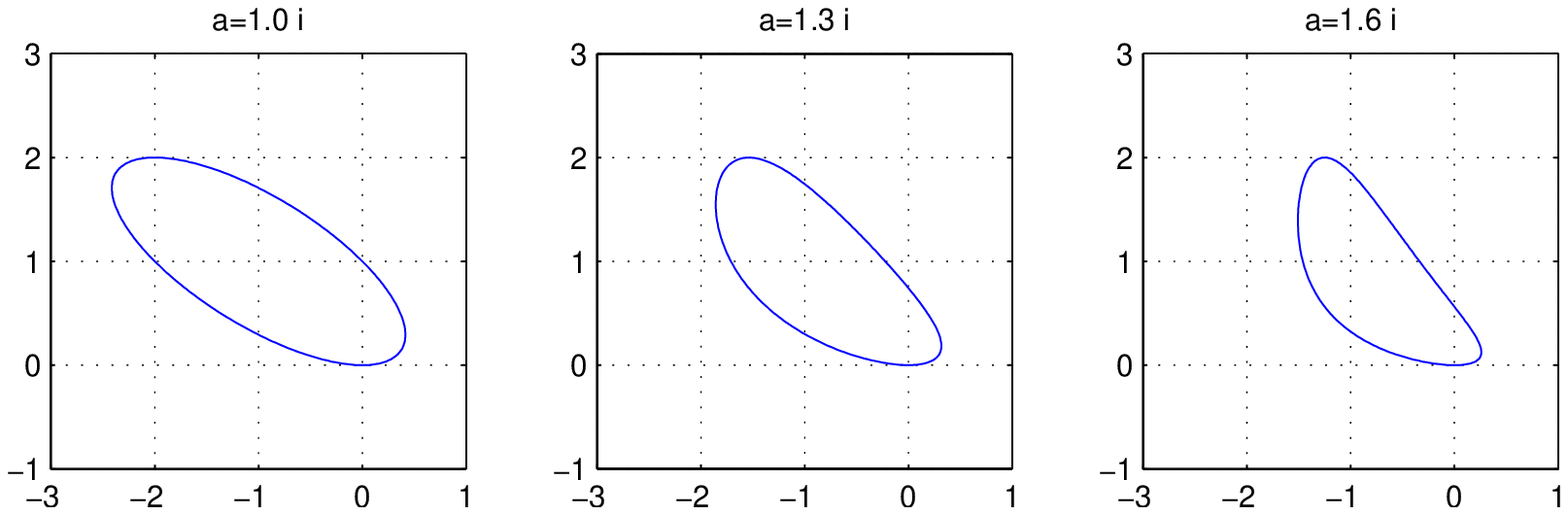,height=4cm} \caption{Various shapes of E$_\Om$-inclusions when varying Im $a$ with Re $a=0$ and $b=1+i$.}\label{varIma}
\end{center}
\end{figure}

Let us now characterize those rational functions $F$ which yield ellipses as E$_\Om$-inclusions. Because $y=u$ on the slit $[-1,1]$, the ellipse takes the shape like the first figure in Figure \ref{varIma} (after translation).
Let the ellipse be given by $x^2+\Ga y^2 + \Gb x y=c$ with $4\Ga > \Gb^2$. Solving for $x$ we get
 \beq
 x= \frac{-\Gb y \pm \sqrt{\Gb^2 y^2 - 4(\Ga y^2-c)}}{2}.
 \eeq{CM9}
Since the discriminant vanishes at $y=\pm 1$, we have $c=4\Ga-\Gb^2$, and hence
 \beq
 x= \frac{-\Gb}{2} y \pm \frac{(1+y)}{2} \sqrt{(4\Ga -\Gb^2) \frac{1-y}{1+y}}.
 \eeq{CM10}
Letting $\zeta= \sqrt{\frac{1-y}{1+y}}$, we have
 \beq
 x= \frac{\pm  \zeta\sqrt{4\Ga -\Gb^2} - \Gb}{\zeta^2+1} + \frac{\Gb}{2} = F(\zeta)
 \eeq{CM11}
for real $\zeta$. It means that ellipses are obtained by $F$'s of the form
 \beq
 F(\zeta)= \frac{b}{\zeta-a} + \frac{b^*}{\zeta-a^*} + c
 \eeq{CM12}
with $a=i$ and $b$ with positive real part.

\medskip
\noindent{\bf Example}. In this example, we construct some E$_\Om$-inclusions other than ellipses. We use $F$ in the form \eq{CM12} with $c=0$ (it amounts to translating the figure). Then in $\zeta$-coordinates $f$ is given by
 \beq
 f(\zeta)= \frac{2i}{\zeta^2+1} + \frac{b}{\zeta-a} + \frac{b^*}{\zeta-a^*}.
 \eeq{CM13}
where both \eq{CM8a} and the absence of real non-zero roots of \eq{CM8b} will be ensured if we choose  $b$ and $-b/a^2$ with positive real parts.

We will plot the image of a vicinity of the real axis in the upper half plane under the map $f$. To avoid computational difficulty in dealing with an infinite space, we use a bilinear transform
 \beq
 \zeta= \frac{1-iw}{w-i},
 \eeq{CM14}
which maps the unit disk onto the upper half plane. Then we plot
 \beq
 f(w)= \frac{2i}{\zeta(w)^2+1} + \frac{b}{\zeta(w)-a} + \frac{b^*}{\zeta(w)-a^*}
 \eeq{CM15}
for $w=r e^{i\theta}$ with $1-\ep \le r \le 1$. From the expansions for $F(\zeta)$ in powers of $\zeta$ and $1/\zeta$ we see 
that near the bottom and top of the inclusion the boundary is given by
\beq x\approx \Real(b)\sqrt{2(1+y)}+O(1+y),\quad\quad
     x \approx -2\Real(b/a)-\Real(b/a^2)\sqrt{2(1-y)}+O(1-y).
\eeq{CM16}
Thus the bottom and top are positioned at $x=0$ and $x=-2\Real(b/a)$ and the curvature of the boundary there is determined by $\Real(b)$ and  
$-\Real(b/a^2)$ respectively.

\begin{figure}[htbp]
\begin{center}
\epsfig{figure=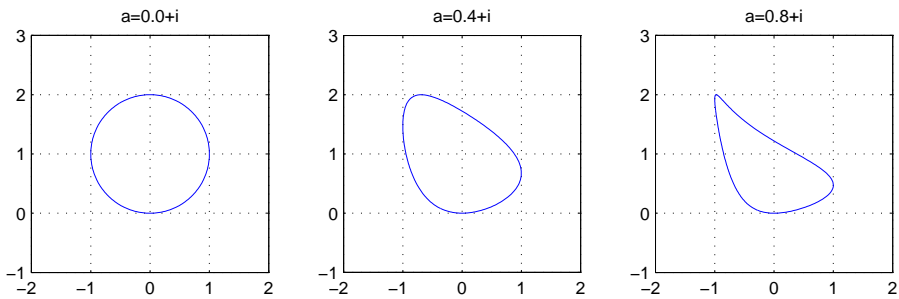,height=4cm} \caption{Various shapes of E$_\Om$-inclusions when varying Re $a$ with  Im $a=1$ and $b=1$.} \label{varRea}
\end{center}
\end{figure}

\begin{figure}[htbp]
\begin{center}
\epsfig{figure=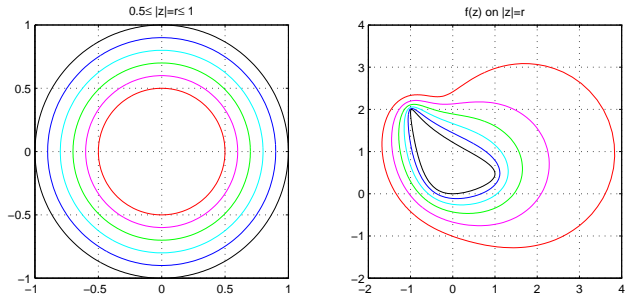,height=4cm} \caption{With $a=0.8+i$ and $b=1$, the inner most curve (the image of $|z|=1$) is the boundary of the E$_\Om$-inclusion (the rightmost inclusion in Fig. \ref{varRea}). The others are images of $|z|=0.9, \ 0,8, \ 0,7, \ 0,6, \ 0,5$. These, or any simple closed curve enclosed by them, can be regarded as boundaries of $\Om$.} \label{varRad}
\end{center}
\end{figure}

\begin{figure}[htbp]
\begin{center}
\epsfig{figure=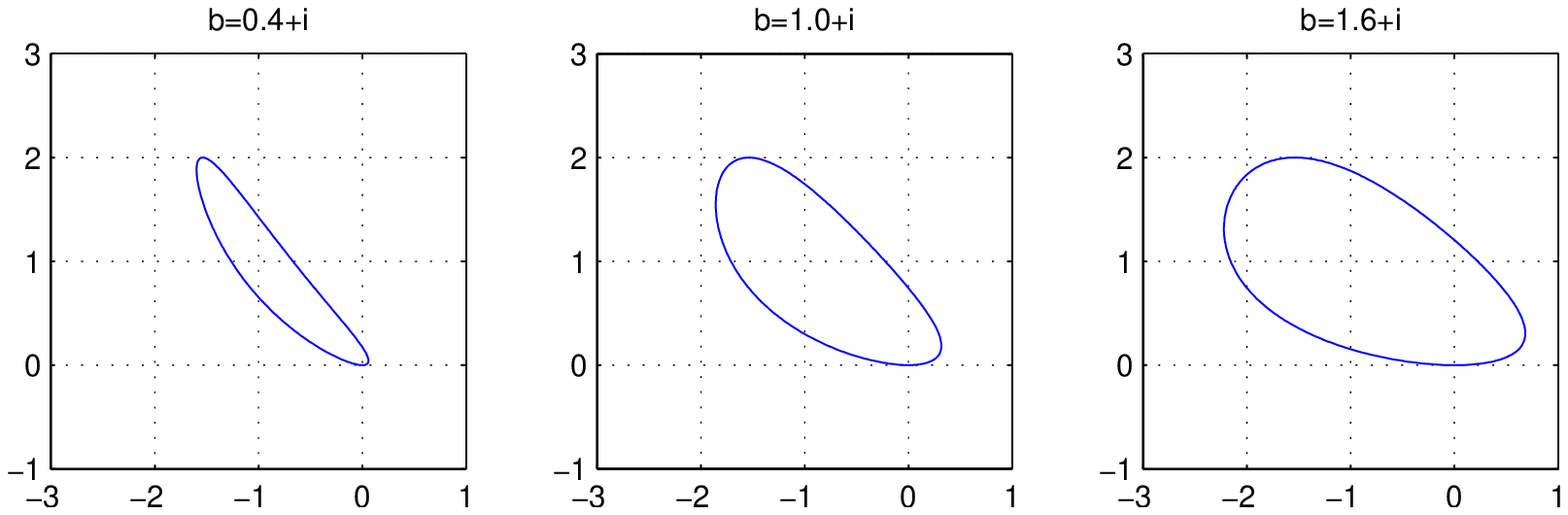,height=4cm} \caption{Various shapes of E$_\Om$-inclusions when varying Re $b$ with Im $b=1$ and $a=1.3i$.}\label{varReb}
\end{center}
\end{figure}

\begin{figure}[htbp]
\begin{center}
\epsfig{figure=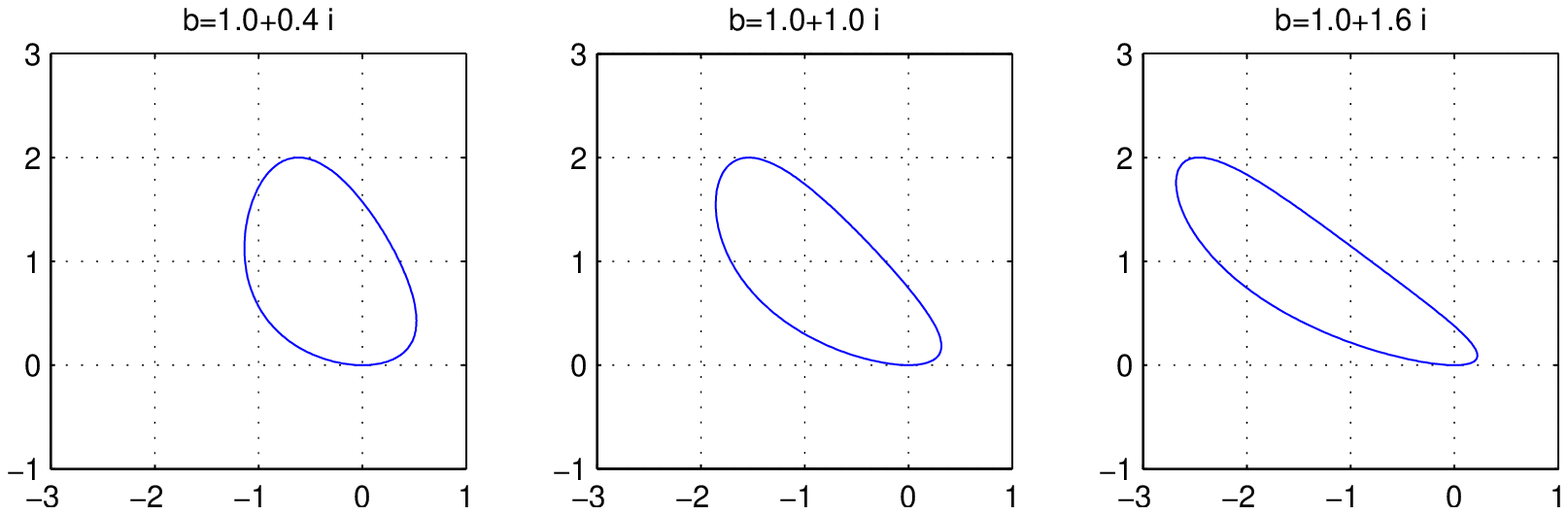,height=4cm} \caption{Various shapes of E$_\Om$-inclusions when varying Im $b$ with Re $b=1$ and $a=1.3i$.}\label{varImb}
\end{center}
\end{figure}

\begin{figure}[htbp]
\begin{center}
\epsfig{figure=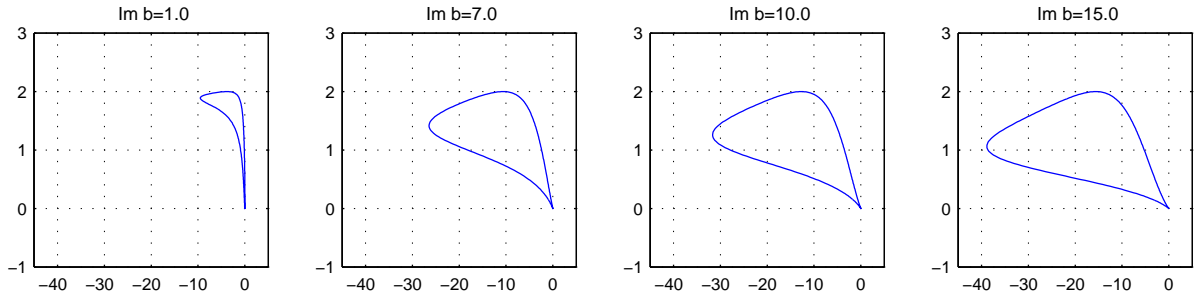,height=4cm} \caption{Various shapes of E$_\Om$-inclusions when varying Im $b$ with Re $b=0.1$ and $b/a^2=-10+2i$.}\label{varImb2}
\end{center}
\end{figure}

\begin{figure}[htbp]
\begin{center}
\epsfig{figure=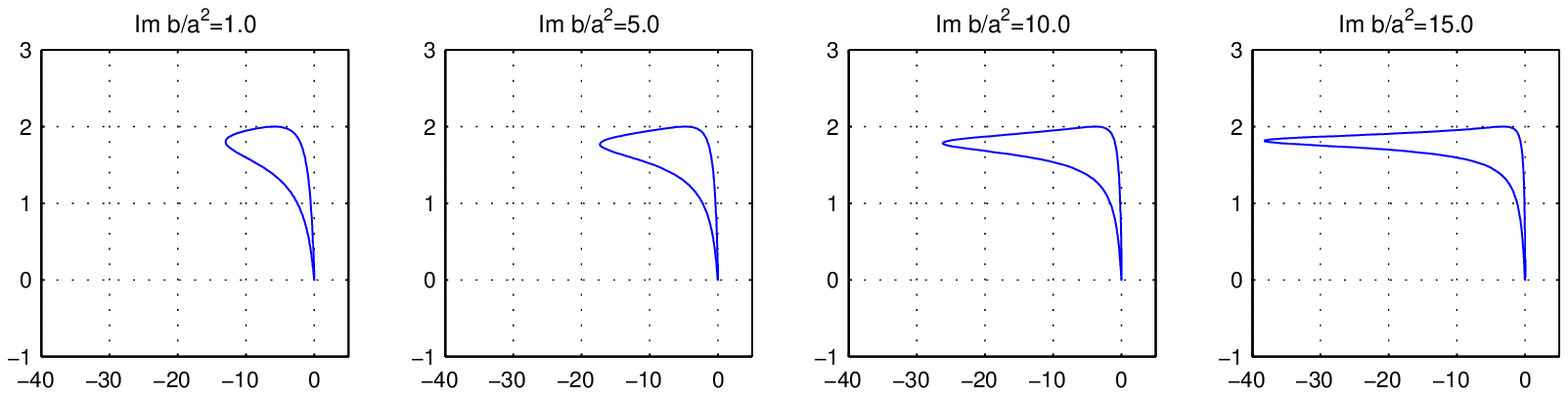,height=4cm} \caption{Various shapes of E$_\Om$-inclusions when varying Im $b/a^2$ with $b=0.1+2i$ and Re $b/a^2=-10$.}\label{varImba^2}
\end{center}
\end{figure}

Figure \ref{varRad} shows various shapes of $\p\Om$, which are the image of $|z|=r<1$ under $f$, and the boundary of E$_\Om$-inclusion, which is the image of $|z|=1$. Figure \ref{varRea}, \ref{varIma}, \ref{varReb}, \ref{varImb}, \ref{varImb2}, and \ref{varImba^2} show various shapes of E$_\Om$-inclusions when we vary the complex parameters $a$, $b$, and $b/a^2$.

We emphasize that with these values of $a$ and $b$, the univalence of $f$ is guaranteed by Lemma \ref{lemma21}.

\section*{Acknowledgements}
The authors thank Michael Vogelius for comments on a draft of the manuscript, and for spurring the interest of GWM in this problem through a lecture at the Mathematical Sciences
Research Institute. GWM is grateful for support from the Mathematical Sciences
Research Institute and from National Science Foundation through
grant DMS-0707978. HK is grateful for support from National Research Foundation through grants No. 2009-0090250 and 2010-0017532, and from Inha University. The work of EK was supported by Korea Research
Foundation, KRF-2008-359-C00004.


\end{document}